\documentclass{article}

\usepackage[utf8]{inputenc}
\usepackage{comment}
\usepackage{tikz}

\usepackage{fullpage}
\usepackage{graphicx}
\usepackage{amsmath}
\usepackage{amssymb}
\usepackage[version=4]{mhchem}
\usepackage{siunitx}
\usepackage{tabularx}
\usepackage{multirow}
\usepackage{longtable,tabularx}
\usepackage{multicol}
\usepackage{circuitikz}
\usepackage{float}
\usepackage{placeins}
\usepackage{booktabs}
\usepackage{cite}
\usepackage[normalem]{ulem} 

\usepackage{xcolor}
\usepackage{soul}

\DeclareMathOperator*{\argmin}{arg\,min}

\def \d {\mathrm{d}}

\def \pp#1#2{\frac{\partial #1}{\partial #2}}

\newcommand{\bx}{\mathbf{x}}
\newcommand{\bu}{\mathbf{u}}
\newcommand{\bc}{\mathbf{c}}

\newcommand{\Dpartial}[2]{\frac{\partial #1}{\partial #2}}

\def \Kn {\mathrm{Kn}}

\newcommand{\bq}{{\mathbf{q}}}

\newcommand{\bF}{{\mathbf{F}}}
\newcommand{\bU}{{\mathbf{U}}}
\newcommand{\bV}{{\mathbf{V}}}

\title{Optimization of Second-Order Transport Models for Transition-Continuum Flows}

\author{Mikolaj Kryger\thanks{Corresponding  author, mkryger@nd.edu} \and Jonathan F. MacArt}
\date{{\normalsize\textit{Department of Aerospace and Mechanical Engineering} \\
    \textit{University of Notre Dame, Notre Dame, Indiana, 46556, USA}}}

\begin{document}

\maketitle

\begin{abstract}
Modeling transition-continuum hypersonic flows poses significant challenges due to thermodynamic nonequilibrium and the associated breakdown of the continuum assumption. Standard continuum models such as the Navier--Stokes equations are inaccurate for these flows, and molecular models can be inefficient due to the large number of computational particles required at moderately high densities. We explore computational modeling of transition-continuum flows using a second-order constitutive theory that provides closures for the terms representing the molecular transport of momentum and energy. We optimize the second-order model parameters for one-dimensional viscous shocks using an adjoint-based optimization method, with the objective function comprising the primitive flow variables. Target data is obtained from moments of distribution functions obtained by solving the Boltzmann equation. We compare results using optimized second-order models, the unoptimized second-order model, and the first-order Navier--Stokes model for Mach numbers $M\in[1.1,10]$ and observe improvements to the shock profiles and shock thickness calculations. We validate the optimized models by comparing the predicted viscous stress and heat flux, which are not included in the objective function, to those obtained by integrating {the} distribution function. The close match to these moments indicates that the satisfactory performance of the optimized second-order models is consistent with the nonequilibrium flow physics.
\end{abstract}

\section{Introduction}\label{sec:Intro}
The physics of high-enthalpy, hypersonic, nonequilibrium flows such as those encountered during atmospheric re-entry or sustained hypersonic high-altitude flight require accurate modeling for engineering predictions. However, the typical Navier--Stokes equations are inadequate in this regime due to the flow's deviation from continuum gas dynamics. This determination can be made based on the Knudsen number, $\Kn = \lambda/L$, where $\lambda$ is the molecular mean free path and $L$ is a characteristic length scale of the flow. The continuum assumption is generally considered valid for $\Kn \leq 0.01$ \cite{anderson2003modern}, while flows at Knudsen numbers above this threshold diverge increasingly from local thermodynamic equilibrium due to the reduced frequency of molecular collisions. This diminishes the accuracy of the linear Fourier's and Newton's laws of heat and momentum transport. The subcontinuum regimes of Knudsen number can be further divided into the $0.01 < \Kn < 10$ transition-continuum regime and the $10 \leq \Kn$ free-molecule regime.

The statistical properties of nonequilibrium flows can be accurately represented using the Boltzmann equation, the standard form of which describes the evolution of the molecular velocity distribution function in physical space, velocity space, and time. Direct simulation Monte Carlo (DSMC) is one of the more prevalent methods of solving the Boltzmann equation~\cite{Cercignani1994} and can be computationally tractable for the mid- to upper-transitional and free-molecule regimes~\cite{SCHWARTZENTRUBER201566, Boyd_Schwartzentruber_2017}.
However, DSMC becomes unwieldy for increasing particle number densities, as are required for lower-Knudsen number flows, and when additional physical and/or velocity space dimensions are needed, for example, in turbulent and/or reacting flows. Accurate, reduced-order equations are desirable for these scenarios.

The Boltzmann equation can be approximated to decrease its computational cost.
The moment approach replaces the solution of the velocity distribution function with transport equations for the higher moments \cite{grad1949kinetic,grad1952profile,muller1998rational}; however, some moment equation-based methods suffer limitations including violations of the second law of thermodynamics~\cite{torrilhon2016modeling}. 
To address this, a class of generalized hydrodynamics models has been developed that satisfies the entropy constraint~\cite{eu1980modified, al1997generalized, myong1999thermodynamically}. Such thermodynamically consistent moment methods have been successfully applied to rarefied one-dimensional shocks and, more recently, external hypersonic flows with considerable improvements over the standard Navier--Stokes equations~\cite{jiang2019computation}. 

The cost of solving the Boltzmann equation or its moments can be further alleviated using the Chapman--Enskog technique, which applies a series expansion in $\Kn$ and assumes the Maxwell--Boltzmann distribution in velocity space~\cite{vincenti1965introduction, BELLOMO199521}. The Navier--Stokes equations are obtained from the series expansion truncated to the first order in $\Kn$~\cite{BELLOMO199521}. Although they are computationally efficient to solve, the first-order equations require continuum fluids in thermodynamic equilibrium, which results in poor accuracy for flows far from these conditions (e.g., transition-continuum flows). The Burnett equations, obtained by including second-order Chapman--Enskog terms, are potentially more accurate for transition-continuum flows than the Navier--Stokes equations~\cite{agarwal2001beyond} but suffer from several difficulties including the need to provide higher-order boundary conditions~\cite{1994AIAAJ..32Q.985L} and potential violations of the second law of thermodynamics~\cite{comeaux1995analysis}. In particular, the entropy production rate predicted by the Burnett equations can be negative if the local Knudsen number and Mach number ($M=u/a$, with $u$ the local bulk velocity magnitude and $a$ the local sound speed) satisfy the relationship $\Kn > 0.4944/M$ in the special case of monatomic gas expansion in the direction of the flow~\cite{comeaux1995analysis}. This may occur in regions of deviation from local thermodynamic equilibrium, including shock waves and boundary layers. More recently, variants of the Burnett equations have been derived that satisfy the second law of thermodynamics as well as the Onsager symmetry principle~\cite{Jadhav_Gavasane_Agrawal_2021}.

Alternatives to Newton's viscosity law and Fourier's heat conduction law have been proposed to extend the validity of the Navier--Stokes equations into the transition-continuum regime. Paolucci \& Paolucci~\cite{paolucci2018second} and Paolucci \cite{Paolucci-2022} derived an extended form of the heat flux and viscous stress constitutive models that augments the first-order models with terms that contain second-order, nonlinear products of flow-field gradients. A different set of constitutive equations that assumed the stress tensor and heat flux vector to be linearly coupled was investigated by Rana, Gupta \& Struchtrup~\cite{doi:10.1098/rspa.2018.0323}.
Both quantities depend on first-order spatial derivatives of temperature and pressure, resulting in a final form similar to the second-order constitutive formulation of Paolucci \& Paolucci~\cite{paolucci2018second} with the  addition of a linear coupling term. Brenner~\cite{BRENNER200511, GREENSHIELDS_REESE_2007} proposed a constitutive equation for conservation of  the fluid specific volume, which is in general a nonconserved property. Its conservation equation assumes a convection--diffusion--production form, and the heat fluxes and viscous stresses derived from this volume transport equation depend on the local density gradients through constitutive functionals for the specific volume diffusive flux and production rate. 

The principal difficulty of applying these second-order models consists in obtaining trustworthy values of their higher-order transport coefficients, each of which would represent a considerable challenge to measure experimentally or to obtain from the distribution function. These transport coefficients have typically been modeled using (i) the known values of similar properties \cite{BRENNER200511} or (ii) power-law closures with manually adjusted parameters~\cite{paolucci2018second}. Since accurate data for the higher-order transport coefficients are unavailable, their model parameters must be optimized based on the accuracy of the flow solution itself compared to computable moments of the distribution function (e.g., density, velocity, heat flux, and viscous stress). To address these challenges, we propose a robust method to obtain these model parameters by optimizing over the parameterized Navier--Stokes partial differential equations (PDEs) including the second-order transport terms.

Most machine learning methods for PDEs have focused on optimizing model parameters to trusted data in an \emph{a priori} sense, that is, before substituting the optimized model into the PDE system to make \emph{a posteriori} predictions. For example, given filtered direct numerical simulation (DNS) data for a turbulent flow, the parameters of a turbulence model could be optimized \emph{a priori} to predict the DNS-evaluated subgrid stresses. The challenges with this approach are twofold. First, parameter optimization does not commute with solving a nonlinear PDE: for example, the model inputs produced by a predictive large-eddy simulation (LES) would have different physical and numerical errors than the filtered DNS inputs used for \emph{a priori} optimization. Thus, the accuracy and even stability of an \emph{a priori}-trained model are not guaranteed when it is used for \emph{a posteriori} predictions~\cite{SIRIGNANO2020109811}. Second, and potentially more problematic for transition-continuum modeling, \emph{a priori} optimization requires trusted data for the closure model outputs. This is difficult or impossible in cases of ``unknown physics,'' for which the target quantities---here, second-order transport coefficients---cannot be directly evaluated from the trusted data (e.g., the Boltzmann distribution function)~\cite{nair2023deep}.

The challenges with \emph{a priori} optimization have motivated the development of optimization methods that target the PDE solution directly and thus embed the optimization procedure within the prediction---\emph{a posteriori} optimization. Deep reinforcement learning (DRL) \cite{li2017deep} has emerged as tool for \emph{a posteriori} optimization with applications to flow modeling \cite{rabault2019artificial,garnier2021review}; however, DRL suffers high optimization costs due to the need to  approximate the PDE system using a trained deep neural network \cite{Liu2024}. Adjoint-based optimization instead optimizes directly over the original PDE system by solving a system of adjoint PDEs, the solution of which provides the gradients needed for parameter optimization \cite{SIRIGNANO2020109811, Duraisamy2021}. The cost of solving the adjoint PDEs is comparable to that of solving the forward PDE system, thus the method is computationally efficient even for high-dimensional parameter spaces. Numerous applications to LES \cite{SIRIGNANO2020109811,PhysRevFluids.6.050502,Sirignano_MacArt_2023} and Reynolds-averaged Navier--Stokes (RANS) \cite{Sirignano2023a} turbulence modeling, as well as flow control \cite{Liu2024} and first-order transition-continuum modeling \cite{nair2023deep}, have demonstrated the utility of adjoint-based optimization for fluid dynamics.

This work presents an application of the adjoint-based optimization approach to the second-order continuum theory of Paolucci \& Paolucci~\cite{paolucci2018second} for transition-continuum flow. We apply the method to canonical one-dimensional (1D), nonequilibrium, transition-continuum viscous shocks in argon \cite{Alsmeyer_1976} for upstream Mach numbers 1.1 to 10. We also compare the performance of the adjoint-based approach to standard, \textit{a priori} optimization for accuracy and stability of the flow predictions.
Section~\ref{sec:Math} introduces the governing equations, second-order continuum closure, and 1D shock problem. Section~\ref{sec:Optim} provides the optimization methods, loss function definitions, and distributed training scheme, and compares \textit{a priori} to \textit{a posteriori} optimization. Section~\ref{sec:Results} analyzes the accuracy and performance of the adjoint-optimized second-order models. Section~\ref{sec:Summary} presents conclusions and future research.

\section{Model Mathematical Background, Methodology and Test Problem} \label{sec:Math}
\subsection{Governing Equations}

We model continuum fluid dynamics using the single-component, compressible Navier--Stokes equations.
Let $\mathbf{U} = \left[ \rho, \rho \mathbf{u}, \rho E \right]^\top$ be the vector of conserved quantities, where $\rho$ is the mass density, $\mathbf{u} \in \mathbb{R}^{N_d}$ is the velocity vector, $N_d$ is the number of space dimensions, and $E = e + \mathbf{u}^\top \mathbf{u}/2$ is the total energy, where $e$ is the specific internal energy. We consider calorically perfect gases, for which $e = c_v T$, $c_v$ is the specific heat at constant volume, and $T$ is the sum of the translational, rotational, and vibrational temperatures. The ideal gas equation of state is $p = \rho R T$, where $p$ is the pressure and $R$ is the specific gas constant.
With a slight abuse of notation, the balance of the conserved quantities in the time-steady case is
\begin{align}
    \Dpartial{\mathbf{U}}{t} = \mathbf{F}\left( \mathbf{U}; \theta\right) = \nabla \cdot \left[ -\mathbf{f}_c\left( \mathbf{U}\right) + \mathbf{f}_d\left( \mathbf{U}; \theta\right) \right] = 0, \label{eq:governing}
\end{align}
where $\theta\in\mathbb{R}^{N_\theta}$ are model parameters requiring calibration, $N_\theta$ is the parameter space dimensionality, and 
\begin{align}
    \mathbf{f}_c\left(\mathbf{U}\right) = 
    \begin{bmatrix}
        \rho \mathbf{u}^\top\\
        \rho \mathbf{u} \mathbf{u}^\top \\ 
        \rho E \mathbf{u}^\top
    \end{bmatrix}
    \quad\mathrm{and}\quad
    \mathbf{f}_d\left( \mathbf{U}; \theta \right) = 
    \begin{bmatrix}
        0\\
        \sigma \left( \mathbf{U}; \theta \right) \\
        \sigma \left( \mathbf{U}; \theta \right) \cdot \mathbf{u} - \mathbf{q}\left( \mathbf{U}; \theta \right)
    \end{bmatrix}
\end{align}
are the conserved and diffusive flux vectors, where 
$\sigma$ is the Cauchy stress tensor and $\bq$ is the heat flux. We map the conserved quantities to a (nonunique) set of primitive quantities $\mathbf{U} \mapsto \mathbf{V}$:
\begin{align}
\mathbf{U} = \left\{
\begin{array}{ll}
    \rho, \\
    \rho \mathbf{u}, \\
    \rho E,
\end{array}
\right. \quad\quad
\mathbf{V} = \left\{
\begin{array}{ll}
    \rho, \\
    \mathbf{u} = \frac{\rho \mathbf{u}}{\rho}, \\
    T = \frac{\left(\gamma-1\right)}{\rho R}\left(\rho E - \frac{\rho\left(\mathbf{u}^\top \mathbf{u}\right)}{2}\right), \label{eq:UtoVmap}
\end{array}
\right.
\end{align}
where $\gamma$ is the specific heat ratio. Utilizing this mapping, the flux divergence can be represented as either $\bF(\bU;\theta)$ or $\bF(\bV;\theta)$, which  will be used in the construction of the adjoint equations (Section~\ref{sec:adjoint}).

\subsection{Second-order Constitutive Equations}\label{sec:Constitutive_Equations}

The PDE system needs to be closed by a set of constitutive equations for the heat flux and the  stress tensor.
These are obtained by proposing a set of constitutive quantities that describe the response functionals of the fluid.
Applying the standard principles of continuum mechanics, the proposed functionals yield a complete material description that obeys the governing conservation equations and the Clausius--Duhem inequality.
This section briefly presents the derivation of the second-order constitutive theory of fluids following Paolucci \& Paolucci~\cite{paolucci2018second}. We introduce several symbol changes for notational consistency with the present work. 

Paolucci \& Paolucci~\cite{paolucci2018second} proposed a set of constitutive equations with assumed dependence on flow field variables and their gradients. Applying the material frame invariance principle restricts the dependence of closure functionals for an isotropic fluid to the set of constitutive quantities
\begin{align*}
\textit{I} = \{\rho, \nabla \rho, D, T, \nabla T \},
\end{align*}
where $D = \left(\nabla \mathbf{u} +  \left(\nabla \mathbf{u}\right)^\top\right)/2$ is the strain-rate tensor. Constitutive equations for the heat flux and stress tensor functionals may be derived based on this set of dependent quantities, such that
\begin{align*}
\bq &= \hat{\bq}(\rho, \nabla \rho, D, T, \nabla T),\\
\sigma &= \hat{\sigma}(\rho, \nabla \rho, D, T, \nabla T),
\end{align*}
where $\hat{\mathbf{q}}$ is the vector-valued heat flux functional, and $\hat{\sigma}$ is the tensor-valued stress functional. The second law of thermodynamics is enforced using an entropy-restriction analysis based on the Clausius--Duhem inequality. 

A lengthy derivation yields the second-order heat flux and stress tensor functionals~\cite{paolucci2018second}, with repeated indices implying summation:
\begin{align*}
\hat{q}_i &= -\left[\left(k+k^{\prime} D_{(1)}\right) \delta_{i l}+k^{\prime \prime} D_{i l}\right] \nabla_l T-\left(k^{\prime \prime \prime} D_{(1)} \delta_{i l}+k^{\prime \prime \prime \prime} D_{i l}\right) \nabla_l \rho \\
\begin{split}
\hat{\sigma}_{i k} &= \left(-p + \mu_d D_{(1)}-\mu^{\prime} D_{j l} D_{l j}-\frac{k^{\prime}}{T} \nabla_j T \nabla_j T-\frac{k^{\prime \prime \prime}}{T} \nabla_j \rho \nabla_j T\right) \delta_{i k}\ + \\
&\quad \left(2 \mu+\mu^{\prime} D_{(1)}\right) D_{i k}-\frac{k^{\prime \prime}}{T} \nabla_i T \nabla_k T-\frac{1}{2} \frac{k^{\prime \prime \prime \prime}}{T}\left(\nabla_i T \nabla_k \rho+\nabla_i \rho \nabla_k T\right),
\end{split}
\end{align*}
where $\delta_{ij}$ is the Kronecker delta, $k$ is the first-order thermal conductivity, $k^{\prime}$, $k^{\prime\prime}$, $k^{\prime\prime\prime}$, and $k^{\prime\prime\prime\prime}$ are second-order thermal conductivities, $D_{(1)} = d^{(1)} + d^{(2)} + d^{(3)}$ is the first invariant of $D$, where $d^{(i)}$ are the eigenvalues of $D$, $\mu$ is the dynamic viscosity, $\mu_d$ is the dilatational viscosity, and $\mu^{\prime}$ is the second-order viscosity. The Clausius--Duhem inequality requires~\cite{paolucci2018second}
\begin{equation}
    \mu \geqslant 0,\quad\kappa = \mu_d + \frac{2}{3}\mu \geqslant 0,\quad k \geqslant 0 \quad\text{and} \quad\{\mu^{\prime}, k^{\prime}, k^{\prime\prime}, k^{\prime\prime\prime}, k^{\prime\prime\prime\prime}\} \in \mathbb{R}^5,\label{eq:Parameter_Constraints}
\end{equation}
where $\kappa$ {is the bulk viscosity}.
In one space dimension, the second-order constitutive functionals simplify to
\begin{align}
q &= -\left(k + \left(k^{\prime}+k^{\prime \prime}\right)\frac{\partial u}{\partial x}\right) \frac{\partial T}{\partial x} - \left(k^{\prime \prime \prime}+k^{\prime \prime \prime \prime}\right) \frac{\partial u}{\partial x} \frac{\partial \rho}{\partial x} \label{eq:q_dim}, \\
\sigma &= -p+ \left(\frac{4}{3} \mu+\kappa\right) \frac{\partial u}{\partial x} - \frac{k^{\prime}+k^{\prime \prime}}{T}\left(\frac{\partial T}{\partial x}\right)^2-\frac{k^{\prime \prime \prime}+k^{\prime \prime \prime \prime}}{T} \frac{\partial \rho}{\partial x} \frac{\partial T}{\partial x}. \label{eq:sigma_dim}
\end{align}

To generalize the transport coefficient analysis and optimization process, each of the coefficients in~\eqref{eq:q_dim}--\eqref{eq:sigma_dim} can be divided into dimensional and dimensionless factors,
\begin{equation}
\begin{split}
\mu = \mu_{T_0}\mu_{*},\quad\quad \kappa = \mu_{T_0}\kappa_{*},\quad\quad k = k_{T_0}k_{*},\quad\quad\quad\quad\quad\quad\quad\quad\\k^{\prime} = k_{*}^{\prime}\frac{{\mu_{T_0}}^2}{\rho_\infty T_\infty},\quad k^{\prime\prime} = k_{*}^{\prime\prime}\frac{{\mu_{T_0}}^2}{\rho_\infty T_\infty},\quad k^{\prime\prime\prime} = k_{*}^{\prime\prime\prime}\left(\frac{\mu_{T_0}}{\rho_\infty}\right)^2,\quad k^{\prime\prime\prime\prime} = k_{*}^{\prime\prime\prime\prime}\left(\frac{\mu_{T_0}}{\rho_\infty}\right)^2,
\end{split}
\label{eq:dimensionless_coeff}
\end{equation}
where $\mu_{T_0}$ and $k_{T_0}$ are dimensional reference coefficients at a reference temperature, and $\mu_{*}$, $\kappa_{*}$, $k_{*}$, $k^{\prime}_{*}$, $k^{\prime\prime}_{*}$, $k^{\prime\prime\prime}_{*}$, and $k^{\prime\prime\prime\prime}_{*}$ are dimensionless correction factors requiring modeling.
Rewriting the constitutive functionals \eqref{eq:q_dim}--\eqref{eq:sigma_dim} using \eqref{eq:dimensionless_coeff}, the final 1D expressions are obtained:
\begin{align}
q\left( \rho, T, \theta \right) &= -\left(k_{T_0}k_{*} + \left(k^{\prime}_{*}+k^{\prime \prime}_{*}\right) \frac{{\mu_{T_0}}^2}{\rho_\infty T_\infty} \frac{\partial u}{\partial x}\right) \frac{\partial T}{\partial x} - \left(k^{\prime \prime \prime}_{*}+k^{\prime \prime \prime \prime}_{*}\right) \left(\frac{\mu_{T_0}}{\rho_\infty}\right)^2 \frac{\partial u}{\partial x} \frac{\partial \rho}{\partial x} \label{eq:q_x_final}, \\
\sigma\left( \rho, T, \theta \right) &= -p +\mu_{T_0} \left(\frac{4}{3} \mu_{*}+\kappa_{*}\right) \frac{\partial u}{\partial x} - \frac{k^{\prime}_{*}+k^{\prime \prime}_{*}}{T} \frac{{\mu_{T_0}}^2}{\rho_\infty T_\infty}\left(\frac{\partial T}{\partial x}\right)^2-\frac{k^{\prime \prime \prime}_{*}+k^{\prime \prime \prime \prime}_{*}}{T} \left(\frac{\mu_{T_0}}{\rho_\infty}\right)^2 \frac{\partial \rho}{\partial x} \frac{\partial T}{\partial x}. \label{eq:sigma_xx_final}
\end{align}
We next introduce power-law models for the higher-order correction factors and define the parameter space for optimization.

\subsection{Modeling the Higher-Order Transport Coefficients}
The second-order thermal conductivity correction  factors in~\eqref{eq:q_x_final} and \eqref{eq:sigma_xx_final} can be simplified for 1D problems as 
\begin{align*}
    k^{\star}\left( \rho, T, \theta \right) &= k_{*}^{\prime}\left( \rho, T, \theta \right) + k_{*}^{\prime \prime}\left( \rho, T, \theta \right),\\
    k^{\star \star}\left( \rho, T, \theta \right) &= k_{*}^{\prime \prime \prime}\left( \rho, T, \theta \right) + k_{*}^{\prime \prime \prime \prime}\left( \rho, T, \theta \right),
\end{align*}
which reduces the optimization space.

Paolucci \& Paolucci~\cite{paolucci2018second} proposed a power-law dependence of the dimensionless coefficients $\mu_{*}\left( \rho, T, \theta \right)$, $\kappa_{*}\left( \rho, T, \theta \right)$, $k\left( \rho, T, \theta \right)$, $k^{\star}\left( \rho, T, \theta \right)$, and $k^{\star \star}\left( \rho, T, \theta \right)$ on a dimensionless temperature $T_{*} = T/T_{\infty}$ and density $\rho_{*} = \rho/\rho_{\infty}$, where $T_\infty$ and $\rho_\infty$ are freestream values. 

While the continuum-mechanical constraints require the higher-order transport coefficients $k^\star$ and $k^{\star\star}$ to be functions of density and temperature, they do not limit their function form, which allows for arbitrary parameterization. We consider the power-law representations for consistency with Paolucci \& Paolucci~\cite{paolucci2018second}, though this is neither a unique closure nor necessarily globally optimal. Indeed, a better fit to the target data (here, Boltzmann solutions) could likely be obtained by permitting the higher-order transport coefficients to vary more nonlinearly with the flow state, for example, using neural networks, and/or by training over a wider range of Mach and Knudsen numbers.

Power-law models for the dimensionless coefficients are
\begin{align}
  \begin{split}
    \mu_{*}&=\mu_{*}(\rho, T, \theta)=\rho_{*}^{\beta_1} T_{*}^{\gamma_1},\\
    \kappa_{*}&=\kappa_{*}(\rho, T, \theta)=\kappa_0 \rho_{*}^{\beta_2} T_{*}^{\gamma_2},\\
    k&=k(\rho, T, \theta)=\rho_{*}^{\beta_3} T_{*}^{\gamma_3},\\
    k^{\star}&=k^{\star}(\rho, T, \theta)=k_0^{\star} \rho_{*}^{\beta_4} T_{*}^{\gamma_4},\\
    k^{\star \star}&=k^{\star \star}(\rho, T, \theta)=k_0^{\star \star} \rho_{*}^{\beta_5} T_{*}^{\gamma_5},
  \end{split}
  \label{eq:power_laws}
\end{align}
where $\beta_i$ are density exponents, $\gamma_i$ are temperature exponents, $\kappa_0$ is a reference dimensionless bulk viscosity to be optimized, and $k_0^\star$ and $k_0^{\star\star}$ are reference higher-order conductivities that also require optimization. The resulting parameter space is
\begin{align*}
\theta = \{\kappa_0,  k_0^{\star}, k_0^{\star \star}, \beta_1, \beta_2, \beta_3, \beta_4, \beta_5, \gamma_1, \gamma_2, \gamma_3, \gamma_4, \gamma_5\} \in \mathbb{R}^{N_\theta}, \quad N_\theta = 13.
\end{align*}
These parameters need to be adjusted to correctly model the fluid's {transport} properties. Since values of several dimensionless coefficients in \eqref{eq:power_laws} cannot be directly obtained from experiments or higher-order calculations, an optimization method is required that can minimize the error between the PDE solution (e.g., $\bV$) and trusted data.

While adjoint methods do not suffer increased costs with increasing $N_\theta$, perturbation-based optimization methods (e.g., finite differences) have computational costs that scale with $N_\theta$.  Paolucci \& Paolucci~\cite{paolucci2018second} reduced the parameter space dimensionality $N_\theta$ by applying standard power law dependencies for the first-order coefficients~\cite{10.1063/1.1616556}.
For monatomic gases, the Chapman--Enskog expansion gives $\kappa_0 = 0$, which results in $\beta_2 = 0$ and $\gamma_2 = 0$.
Furthermore, Paolucci \& Paolucci used a standard power-law dependence of $\mu_{*}$, $\beta_1 = 0.5$ and $\gamma_1 = 0.72$, which was originally  obtained from experimental data~\cite{10.1063/1.1616556}. For consistency with their work, we adopt the same division of $\theta$ into parameters not subject to optimization $\theta_{\text{non-optim}}$ and parameters subject to optimization $\theta_{\text{optim}}$:
\begin{align*}
    \theta &= \{\theta_{\text{non-optim}}, \theta_{\mathrm{optim}}\}.
\end{align*}
Optimization of the parameters $\mu_{*}$, $\beta_1$, and $\gamma_1$ would likely result in greater model flexibility, though potentially at the cost of generalizability, as these parameters are typically obtained from experiments. This could result in models that potentially do not relax to the standard first-order closures in continuum flow regions.
To satisfy the Clausius--Duhem inequality, the constraint on the parameters subject to optimization is
\begin{align}
    \theta_{\mathrm{optim}} &= \{k_0^{\star}, k_0^{\star \star}, \beta_3, \beta_4, \beta_5, \gamma_3, \gamma_4, \gamma_5\} \in \mathbb{R}^{N_\theta'}, \quad N_\theta' = 8.\label{eq:theta_optim}
\end{align}
The test problem and the numerical methods used to solve the PDEs are discussed in the following section. The optimization approaches and constraints applied to \eqref{eq:theta_optim} and the training data sets are described in Section~\ref{sec:Optim}.

\subsection{Test Problem and Numerical Methods}

We target canonical 1D viscous shocks in transition-continuum flows of argon with upstream Mach numbers $M_{\infty} \in \{1.1, 2, 3, 4, 5, 6, 7, 8, 9, 10\}$,  for which experimental data are available~\cite{Alsmeyer_1976, 10.1063/1.1711007, doi:10.2514/6.1964-35, Schmidt_1969, doi:10.2514/3.49425}. Each case has upstream conditions $p_{\infty} = 6.667$~Pa and $T_{\infty} = 300$~K. The constant ratio of specific heats is $\gamma = 5/3$ with specific gas constant $R = 208.12~\text{kJ}/\left(\text{kg}\cdot\text{K}\right)$.

\begin{figure}
  \centering
  \includegraphics[width=0.6\linewidth]{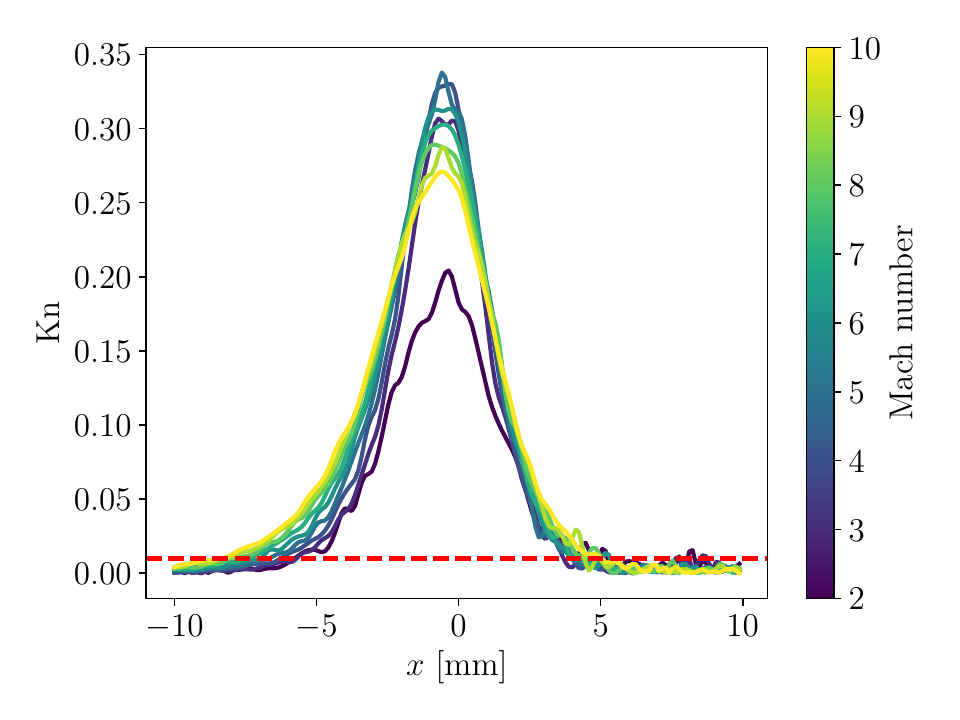}
  \caption{Local Knudsen numbers evaluated from DSMC data  for $M_{\infty} \in \{2, 3, 4, 5, 6, 7, 8, 9, 10\}$. The red line indicates the continuum limit ($\mathrm{Kn}\sim0.01$).}
  \label{fig:Kn_distribution}
\end{figure}

Figure~\ref{fig:Kn_distribution} displays the local density gradient-based Knudsen number
\begin{align*}
    \text{Kn} = \frac{\lambda_\infty}{\rho_{\infty}}\left(\pp{\rho}{x} \right)
\end{align*}
evaluated from the DSMC data (see Section~\ref{sec:Optim}), where $\lambda_\infty$ is the upstream mean free path, for $M_{\infty} \in \{2, 3, 4, 5, 6, 7, 8, 9, 10\}$. The maximum Knudsen number within the shock varies between 0.2 and 0.34, depending on the upstream Mach number, each of which is greater than the continuum limit $\Kn\sim 0.01$. Thus, each of shocks we consider contains regions of substantially transition-continuum flow.

We semidiscretize \eqref{eq:governing} using standard finite-volume schemes. The numerical domain of length $L_x = 30$\,mm, with 20\,mm upstream of the shock, is discretized using $N_x=256$ equispaced elements. Inviscid fluxes are approximated using Roe's flux-difference splitting scheme~\cite{annurev:/content/journals/10.1146/annurev.fl.18.010186.002005}, and viscous fluxes are evaluated to second-order at cell faces. Dirichlet boundary conditions are specified on both the inlet and the outlet, with the outlet values obtained from the Rankine--Hugoniot jump relations.

The  steady-state problem is solved using Newton iteration with step damping and bracketing.
The algorithm can be split into two parts: step calculation and step correction. The Newton step is given by
\begin{align}
  \left.\Dpartial{\mathbf{F}}{\mathbf{U}}\right|_n\Delta \mathbf{U}_n = -\mathbf{F}\left(\mathbf{U}_n\right), \label{eq:Newton_step}
\end{align}
where $\partial{\mathbf{F}}/\partial{\mathbf{U}}$ is the flux Jacobian matrix, and $\Delta \mathbf{U}_n$ is the solution update at iteration $n$. The solution is then updated as $\bU_{n+1}=\bU_n+\Delta\bU_n$. The system \eqref{eq:Newton_step} is solved using Gauss--Jordan elimination routines supplied by the PyTorch library~\cite{NEURIPS2019_9015}.
The iterations are stopped when the relative change of the conserved variables satisfies
\begin{align*}
    \Delta \mathbf{U}_{\text{Rel}} &= \frac{\Delta \mathbf{U}_{n} - \Delta \mathbf{U}_{n-1}}{\mathbf{U}_{\text{Norm}}} \leqslant 10^{-12}, 
\end{align*}
where $\mathbf{U}_{\text{Norm}} = \left[\rho_{\infty}, \:\rho_{\infty}u_{\infty}, \:\rho_{\infty}E_{\infty} \right]^\top$.

Large Newton steps may cause the solution to exceed physical limits by obtaining negative values for the local density and/or pressure. Bracketing is used to adjust the step size to prevent unphysical states and solution overshoots. If the local density and/or pressure of the bracketed solution becomes negative, then the update $\Delta \mathbf{U}$ is multiplied by a damping factor until the resulting $\bU_{n+1}$ remains within physical limits. Full details of the bracketing and damping procedures are provided by Nair \emph{et al.}~\cite{nair2023deep}. The bracketing and damping factors in this work were $0.25$ and $0.9$, respectively.

\section{Optimization Methods} \label{sec:Optim}

We now formulate optimization problems for the reduced parameter space $\theta_\mathrm{optim}$. For brevity, we subsequently denote this reduced set of parameters as $\theta$. The second-order model parameters proposed by Paolucci \& Paolucci~\cite{paolucci2018second} (PP18) are considered in detail in Section~\ref{sec:Results}.

Any optimization process requires high-quality target data. We obtain nonequilibrium velocity distribution functions $f(\bx,\bc)$ for the 1D transition-continuum shocks from DSMC solutions of the Boltzmann equation, where $\bc$ is the independent velocity space of the distribution function.
For \emph{a posteriori} (adjoint-based) optimization, the target values of the primitive variables, $\bV^e$, are obtained as moments  of the distribution function, e.g., for a molecular property $\psi$,
\begin{align}
  \bar{\psi}^e(\bx) = \int_{-\infty}^\infty  \psi f(\bx,\bc)\,\d\bc,
  \label{eq:dist_fn_int}
\end{align}
where we drop the overbar for quantities already appearing in \eqref{eq:governing}. Defining the peculiar velocity $\bc' \equiv \bc - \bu$ and its squared magnitude $c'^2 = \bc'^\top\bc'$, the heat flux and viscous stress needed for \emph{a priori} optimization are obtained as~\cite{vincenti1965introduction}
\begin{align}
    \bq^e &= \frac{1}{2}\rho\overline{\bc' c'^2}, \\
    \tau^e &= -\left(\rho \overline{\bc'\bc'^\top} - p\mathbf{I}\right), \label{eq:Boltzmann_moments}
\end{align}
where $\tau = \sigma + p\mathbf{I}$ is the viscous stress tensor, and $\mathbf{I}\in\mathbb{R}^{N_d\times N_d}$ is the identity matrix.
The target data obtained in this way have been previously validated against the experimental data of Alsmeyer~\cite{Alsmeyer_1976} by Nair \emph{et al.}~\cite{nair2023deep}.

A standard \emph{a priori} optimization approach for the second-order model parameters is described next, and its predictions are compared to those using the PP18 parameters. The adjoint-based, \emph{a posteriori} approach is presented in Section~\ref{sec:a_posteriori}, and its performance is discussed in Section~\ref{sec:Results}.

\subsection{\textit{A priori} Approach}\label{sec:a_priori}

\textit{A priori} optimization comprises the majority of current machine learning approaches. Its advantages include computational simplicity and relative ease of parameter optimization, but it can result in inaccurate and/or unphysical predictions when used in conjunction with PDEs. This is primarily due to its lack of explicit PDE constraints. As a test of this baseline approach, we apply \textit{a priori} optimization to obtain second-order model parameters and evaluate their performance for the 1D shock problem. The following subsections provide the loss function, optimization problem, optimizer settings, and predictive results for \textit{a priori}-optimized parameters.

\subsubsection{Loss Function Definition and Optimization Constraints}

Given a scalar loss (objective) function $J(\theta)$, the \emph{a priori} optimization problem is
\begin{align*}
    &\argmin_\theta\ J(\theta) \\
    &\text{subject to}\ \theta \in \mathbb{R}^{N_\theta'},
\end{align*}
where the constraint is needed to satisfy the entropy inequality \eqref{eq:Parameter_Constraints}. The \emph{a priori} method directly optimizes the constitutive equations' parameters, hence its objective function must be an explicit function of the constitutive terms. We construct the objective function
\begin{equation}
    J(\theta) = \frac{1}{2} \int \Bigg[\frac{1}{\max{\bq^{e}\cdot\bq^e}} \left(\bq(\theta) - \bq^e \right)\cdot\left(\bq(\theta) - \bq^e \right) + \frac{1}{\max{\tau^e:\tau^e}} \left(\tau(\theta) - \tau^e \right):\left(\tau(\theta) - \tau^e \right)\Bigg]\, \d\bx, \label{eq:loss_apriori}
\end{equation}
where the two terms have been normalized by the target data such that their contributions to the objective function are of the same order of magnitude.

\subsubsection{Parameter Optimization} \label{sec:RMSprop_hyper}

Stochastic gradient descent methods update the parameters as
\begin{align}
    \theta_{j+1} = \theta_{j} - \alpha_j \nabla_\theta J, \quad j\in[0,N_\mathrm{iter}]\label{eq:grad_descent},
\end{align}
where $j$ is the optimization iteration number,  $N_\mathrm{iter}$ is the number of optimization iterations, $\alpha_j$ is the learning rate at iteration $j$, and the objective-function gradient $\nabla_{\theta} J$ is calculated by differentiating \eqref{eq:q_x_final}, \eqref{eq:sigma_xx_final}, and \eqref{eq:loss_apriori} using the chain rule. 
We update $\theta_{j+1}\leftarrow\theta_j$ using the RMSprop optimizer provided by the PyTorch library~\cite{NEURIPS2019_9015} with its default settings. 

To increase the stability of the optimization process and to prevent parameter overshoot, the learning rate $\alpha$ is adapted during the optimization process as~\cite{nair2023deep}
\begin{equation}
\alpha_{j+1}=\left\{\begin{array}{cc}
\alpha_j, & \epsilon_{\mathrm{rel}}>\epsilon_{\text{T}, j} \\
0.75 \alpha_j, & \epsilon_{\mathrm{rel}} \leq \epsilon_{\text{T}, j}
\end{array}, \quad \epsilon_{\text{T}, j+1}=\left\{\begin{array}{cc}
\epsilon_{\text{T}, j}, & \epsilon_{\mathrm{rel}}>\epsilon_{\text{T}, j} \\
0.75 \epsilon_{\text{T}, j}, & \epsilon_{\mathrm{rel}} \leq \epsilon_{\text{T}, j}
\end{array}\right.\right.,\label{eq:LR_update}
\end{equation}
where  $\epsilon_{\text{rel}}=J_j/J_0$ is relative error of the loss function at iteration $j$ compared to the initial error $J_0$, and  $\epsilon_{\text{T},j}$ is a relative error threshold. When the relative error decreases below the threshold, \eqref{eq:LR_update} reduces the learning rate to prevent overfitting. For all \emph{a priori} optimization runs, the initial learning rate is $\alpha_0 = 8.0\cdot10^{-3}$, and the initial error threshold is $\epsilon_{\text{T}, 0} = 0.75$.

The flow variables and their gradients, required to evaluate the modeled terms in \eqref{eq:loss_apriori}, are obtained from the DSMC data. We optimize three models individually targeting upstream Mach numbers $M_{\infty} \in\{2, 5, 8\}$. 
Model parameters are initialized at $j=0$ using the PP18 parameters. All models are optimized for $N_\mathrm{iter}=2500$, which results in full, monotonic convergence of the models to a local loss minimum. Due to the reduction of the initial learning rate by over one order of magnitude, no overfitting is observed for models trained using these hyperparameters.

\subsubsection{Optimization Results}

\begin{figure}
    \centering
    \begin{minipage}[t]{0.49\textwidth}
        \centering
        \includegraphics[width=\linewidth]{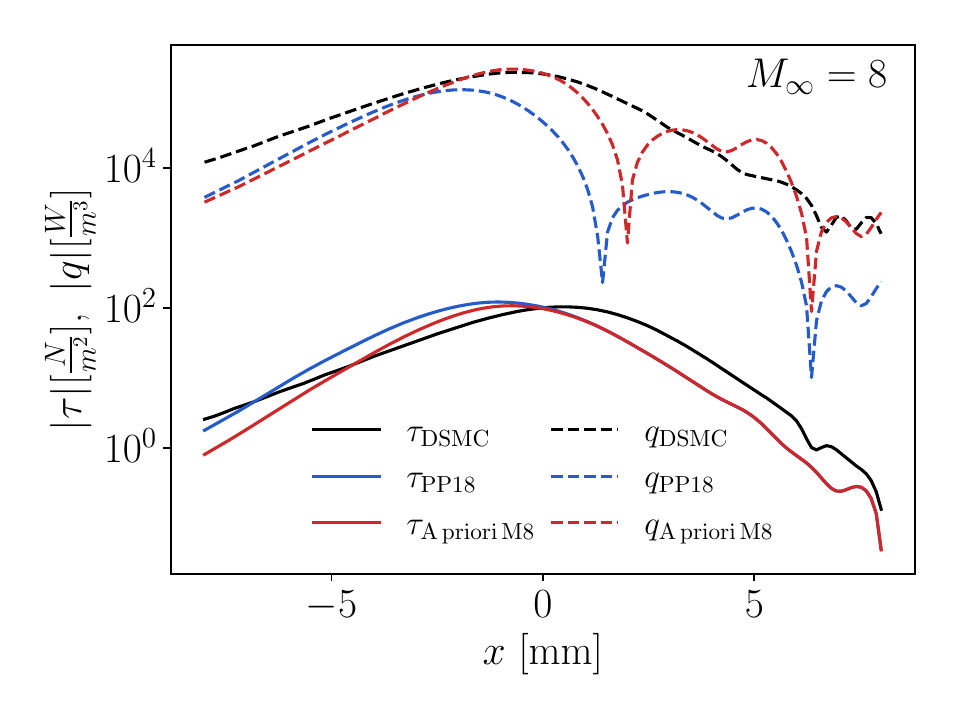}
        \begin{tikzpicture}[remember picture,overlay]
            \node[anchor=north west] at (-3.25,6.25) {\textbf{a)}};
        \end{tikzpicture}
    \end{minipage}
    \hfill
    \begin{minipage}[t]{0.49\textwidth}
        \centering
        \includegraphics[width=\linewidth]{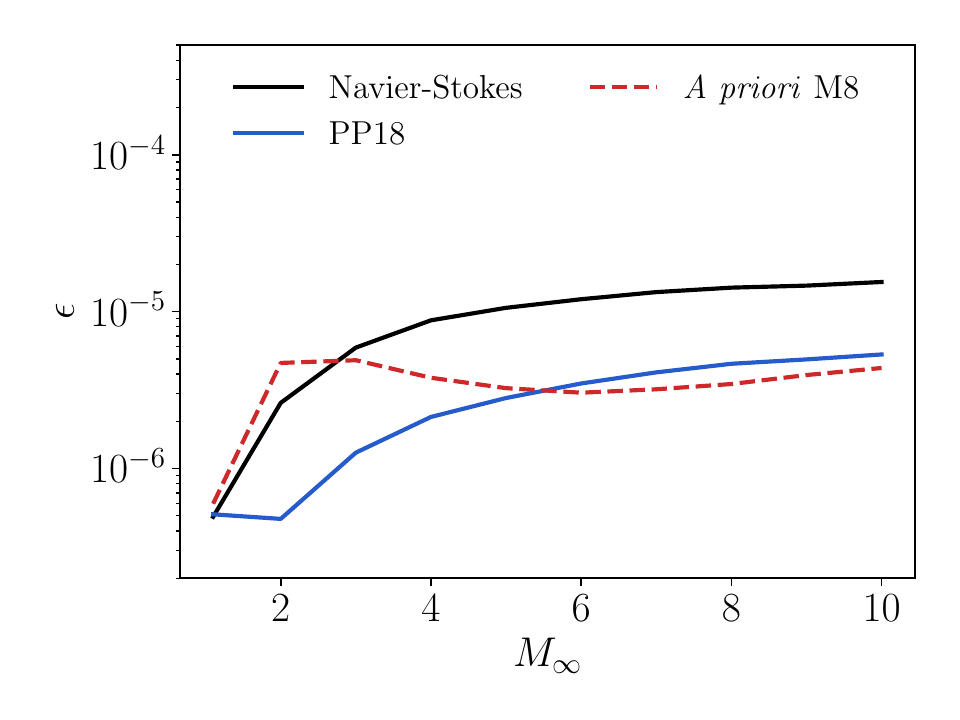}
        \begin{tikzpicture}[remember picture,overlay]
            \node[anchor=north west] at (-3.25,6.25) {\textbf{b)}};
        \end{tikzpicture}
    \end{minipage}
\vspace{-0.5cm}
\caption{(a) In-sample heat flux and {viscous} stress magnitudes for first-order, PP18 second-order, and \textit{a priori} optimized (M8) second-order models. (b) L2 density error of \emph{a posteriori} simulations across the range of testing Mach numbers.}
\label{fig:TauQRhoErr_Apriori}
\end{figure}

Figure~\ref{fig:TauQRhoErr_Apriori}a compares the in-sample, \emph{a priori}, $M_\infty=8$ (M8)-optimized heat flux and viscous stress   to the DSMC-obtained target fields and the unmodified PP18 model predictions. It is apparent that the \emph{a priori}  optimization process  improves these in-sample predictions, particularly upstream of the shock for the viscous stress and downstream of the shock for the heat flux. The \emph{a priori}-trained models for the other Mach numbers produced similarly modified predictions of the viscous stress and heat flux.

However, the out-of-sample performance of the \emph{a priori}-trained second-order models is suboptimal and in some cases worse than the first-order closures. Figure~\ref{fig:TauQRhoErr_Apriori}b compares the \emph{a posteriori} L2 density error across the shock,
\begin{equation}
  \epsilon = \sqrt{\frac{1}{N_x}\sum_{i = 1}^{N_x} \left(\rho^{e}(x_i) - \rho(x_i)\right)^2},
  \label{eq:epsilon}
\end{equation}
for the first-order Navier--Stokes, PP18 second-order, and M8 \emph{a priori}-optimized second-order models evaluated for the entire range of upstream Mach numbers. The optimized M8 model is more accurate than PP18 (and first-order Navier--Stokes) near the in-sample $M_\infty=8$ but is less accurate than PP18 for $M_\infty<5$ and less accurate than the first-order model for $M_\infty<3$. Analogous out-of-sample performance was observed for the other \emph{a priori}-trained models, with some far out-of-sample testing conditions (e.g., the $M_\infty=2$-trained model tested for $M_\infty=10$) resulting in numerical instability. This poor out-of-sample, \emph{a posteriori} performance is the due to the lack of explicit PDE constraints in the \emph{a priori} optimization process~\cite{SIRIGNANO2020109811,PhysRevFluids.6.050502}. We next discuss the \emph{a posteriori} (adjoint-based) optimization procedure; its performance and generalizability are presented in Section~\ref{sec:Results}.

\subsection{\textit{A posteriori} Approach} \label{sec:a_posteriori}

Unlike \emph{a priori} optimization, the \textit{a posteriori} approach applies a PDE constraint to the optimization problem, which aims  to improve the trained models' predictive accuracy  while eliminating the stability issues of  \textit{a priori} optimization. The \emph{a posteriori} optimization problem is
\begin{align*}
    &\argmin_\theta\ J(\bV(\theta)) \\
    &\text{subject to}\ \theta_{\mathrm{optim}} \in \mathbb{R}^{N_\theta'} \quad\wedge \quad\mathbf{F}(\mathbf{V}; \theta) = 0,
\end{align*}
where we emphasize the explicit dependence of the PDE constraint ($\bF=0$) on $\theta$ and the implicit dependence of objective function on $\theta$ via the PDE solution (here, the primitive variables $\bV$). Unlike the \emph{a priori} objective function \eqref{eq:loss_apriori}, the \emph{a posteriori} objective function depends directly on the PDE solution,
\begin{equation}
  J(\bV(\theta)) = \frac{1}{2}\int \bigg[\frac{1}{\bV_\infty}\odot(\bV(\theta) - \bV^e)\cdot(\bV(\theta) - \bV^e)\bigg]\, \d\bx,
  \label{eq:loss}
\end{equation}
where $\odot$ denotes elementwise multiplication, and the primitive variables are chosen over the conserved variables for faster optimization convergence~\cite{nair2023deep}.

\subsubsection{Adjoint-based Optimization} \label{sec:adjoint}

As in the \emph{a priori} approach, the objective-function gradient $\nabla_\theta J$ is needed to update the parameters via stochastic gradient descent. However, given the implicit dependence of \eqref{eq:loss} on $\theta$ via the primitive variables $\bV$, a na\"ive approach yields
\begin{align}
    \nabla_\theta J = \Dpartial{J}{\mathbf{V}} \Dpartial{\mathbf{V}}{\theta}, \label{eq:naive}
\end{align}
where the second factor is high-dimensional ($(N_d+2)\times N_\theta$) and is a function of the PDE constraint. Evaluating the gradient in this manner would require one perturbed PDE solution per element of $\partial\bV/\partial\theta$, which would be computationally intractable for deep learning models, for which $N_\theta$ can be $O(10^5)$ or greater~\cite{nair2023deep}. 
Despite the fact that the optimization space in this work consists of relatively few parameters ($N'_\theta=8$), the adjoint-based approach remains computationally expedient, as fewer than $N'_\theta$ additional adjoint equations are required.

Adjoint sensitivities enable calculation of $\nabla_\theta J$ without the need to evaluate $\partial \mathbf{V}/\partial \theta$ directly. We introduce  adjoint variables $\hat{\mathbf{V}} = [\hat{\rho}, \hat{u}, \hat{T}]^\top$,
with which we can write the Lagrangian for the optimization problem,
\begin{align}
    \mathcal{L}(\mathbf{V}, \theta, \hat{\mathbf{V}}) = J(\mathbf{V}(\theta)) + \hat{\mathbf{V}}^\top \mathbf{F}(\bV;\theta).
\end{align}
For the present steady-state problem, we have $\mathbf{F} = 0$, which holds in the discrete case as long as the PDE solution is sufficiently converged; hence, the Lagrangian and the objective function are equivalent. Taking the gradient with respect to $\theta$, we obtain
\begin{align*}
  \nabla_\theta J = \nabla_{\theta} \mathcal{L} = \Dpartial{J}{\mathbf{V}} \Dpartial{\mathbf{V}}{\theta} + \Dpartial{\hat{\mathbf{V}}^\top}{\theta} \mathbf{F} + \hat{\mathbf{V}}^\top \left(\Dpartial{\mathbf{F}}{\mathbf{V}} \Dpartial{\mathbf{V}}{\theta} + \Dpartial{\mathbf{F}}{\theta}\right),
\end{align*}
which with $\mathbf{F} = 0$ simplifies to
\begin{align}
    \nabla_\theta J = \left( \Dpartial{J}{\mathbf{V}} + \hat{\mathbf{V}}^\top \Dpartial{\mathbf{F}}{\mathbf{V}}\right) \Dpartial{\mathbf{V}}{\theta} + \hat{\mathbf{V}}^\top \Dpartial{\mathbf{F}}{\theta} \label{eq:zero}.
\end{align}
To avoid the need to calculate $\partial{\mathbf{V}}/\partial{\theta}$, we require the adjoints to satisfy
\begin{align}
    \left( \Dpartial{\mathbf{F}}{\mathbf{V}} \right)^\top \hat{\mathbf{V}} = - \left( \Dpartial{J}{\mathbf{V}} \right)^\top,\label{eq:adjoint}
\end{align}
in which we evaluate $\partial{\mathbf{F}}/\partial{\mathbf{V}}$ and $\partial{J}/\partial{\mathbf{V}}$ discretely using algorithmic differentiation over the semidiscretized PDE right-hand side. In the present 1D cases, we solve the adjoint system~\eqref{eq:adjoint} using Gauss--Jordan elimination; higher-dimensional problems would require sparse Jacobian representations and linear solvers.
 Having satisfied the adjoint equation \eqref{eq:adjoint}, we obtain the loss function gradient from \eqref{eq:zero},
\begin{align}
    \nabla_\theta J = \hat{\mathbf{V}}^\top \Dpartial{\mathbf{F}}{\theta}, \label{eq:Final_grad_optim}
\end{align}
which does not require the potentially expensive evaluation of $\partial{\mathbf{V}}/\partial{\theta}$. The resulting gradients are used for gradient-descent parameter updates; see Section~\ref{sec:RMSprop_hyper}.

\subsubsection{Biased Minibatch Training}

We apply parallelized (``minibatch'') optimization to ensure the accuracy of the optimized models across the Mach number range~\cite{nair2023deep}. This is done using message passing interface (MPI) communication to average the gradients $\nabla_\theta J_i$, $i\in[1,N_m]$, where $N_m$ is the number of simultaneous minibatch simulations. Since higher $M_\infty$ shocks have larger $J_0$, unbiased minibatch optimization shows an inherent bias toward these cases. Unbiased minibatch-optimized models are therefore  less accurate at lower $M_\infty$ than at higher $M_\infty$.  Additionally, the flow field gradient magnitudes also increase with $M_\infty$, which makes the PDE system more sensitive to parameter changes at higher $M_\infty$. This results in potentially larger magnitudes of $\nabla_\theta J$ at the upper range of the training $M_\infty$.

We propose a biased minibatch scheme to mitigate this effect.
We define a weighted average of the parameter gradients
\begin{align}
    \nabla_\theta J = \sum_{i=1}^{N_m}\frac{\chi_i}{\sum_{j=1}^{N_m} \chi_j} \nabla_{\theta} J_i, \label{eq:parameter_step}
\end{align}
where $\chi_j$, $j\in[1,N_m]$ are loss gradient biases. A given bias $\chi_j$ is assigned to the MPI process optimizing over $M_{\infty,j}$, with lower-Mach cases receiving larger biases than higher-Mach cases. All minibatch-trained models are optimized over $M_\infty\in\{2, 5, 8\}$; three sets of biases are explored: $\chi=[1,1,1]$ (unbiased), $\chi=[3,1,1]$ and $\chi=[7,2,1]$. The choice of biases was not exhaustively investigated in search of their optimum values. The chosen sets serve to investigate the effect of increased contribution of the lower Mach number cases. Magnifying the influence of these cases aims to counteract the inherent bias of the loss function toward the higher Mach number cases.

\subsubsection{Initial parameters and early stopping}

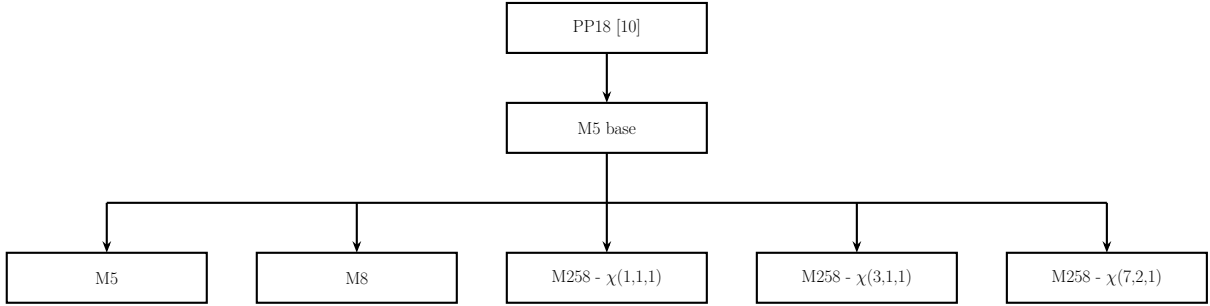
\begin{figure}
\centering
\resizebox{0.97\textwidth}{!}{%
\begin{circuitikz}
\tikzstyle{every node}=[font=\Huge]
\node [font=\LARGE] at (-5,9.5) {};
\draw [ line width=3pt ] (6.25,18.25) rectangle  node {\Huge PP18~\cite{paolucci2018second}} (16.25,15.75);
\draw [line width=3pt, ->, >=Stealth] (11.25,15.75) -- (11.25,13.25);
\draw [, line width=3pt ] (6.25,13.25) rectangle  node {\Huge M5 base} (16.25,10.75);
\draw [line width=3pt, ->, >=Stealth] (11.25,10.75) -- (11.25,5.75);
\draw [, line width=3pt ] (6.25,5.75) rectangle  node {\Huge M258~-~$\chi$(1,1,1)} (16.25,3.25);
\draw [, line width=3pt ] (-6.25,5.75) rectangle  node {\Huge M8} (3.75,3.25);
\draw [, line width=3pt ] (18.75,5.75) rectangle  node {\Huge M258~-~$\chi$(3,1,1)} (28.75,3.25);
\draw [, line width=3pt ] (-18.75,5.75) rectangle  node {\Huge M5} (-8.75,3.25);
\draw [, line width=3pt ] (31.25,5.75) rectangle  node {\Huge M258~-~$\chi$(7,2,1)} (41.25,3.25);
\draw [line width=3pt, short] (-13.75,8.25) -- (36.25,8.25);
\draw [line width=3pt, ->, >=Stealth] (-13.75,8.25) -- (-13.75,5.75);
\draw [line width=3pt, ->, >=Stealth] (-1.25,8.25) -- (-1.25,5.75);
\draw [line width=3pt, ->, >=Stealth] (23.75,8.25) -- (23.75,5.75);
\draw [line width=3pt, ->, >=Stealth] (36.25,8.25) -- (36.25,5.75);
\end{circuitikz}
}%
\caption{Model training hierarchy for \textit{a posteriori} optimization.}
\label{fig:ModelTrainingStructure}
\end{figure}

The stability of the \textit{a posteriori} optimization procedure depends on the initial parameters $\theta_0$ due to the optimization procedure's close coupling to the PDE solution. The PP18 parameters ($\theta_\mathrm{PP18}$) are accurate for $M_\infty\in[1.1,2]$~\cite{paolucci2018second} but are less accurate at higher $M_\infty$. For example, adjoint-based optimization at $M_{\infty} = 8$ of a model initialized with $\theta_\mathrm{PP18}$ showed poor numerical stability during training. To improve the stability of the optimization process over $M_\infty\in[1.1, 10]$, we first optimize the PP18 parameters for a single case, $M_\infty=5$, over $N_\mathrm{iter}=1200$ iterations with initial learning rate $\alpha_0 = 8.0 \cdot 10^{-3}$. This base model (``M5 base'') provides the initial parameters for all subsequently optimized models. We then resume training for $M_\infty=5$ (M5), $M_\infty = 8$ (M8), and the minibatch-trained models (M258); Figure~\ref{fig:ModelTrainingStructure} illustrates this model training hierarchy.
The initial learning rates of these subsequent models were adjusted manually to ensure that each model converges within $N_\mathrm{iter}=1200$; these were $\alpha_0 = 2.0\cdot 10^{-3}$ for M5, M8, and M258~-~$\chi$(1,1,1) and $\alpha_0 = 4.0\cdot 10^{-1}$ for  M258~-~$\chi$(3,1,1) and M258~-~$\chi$(7,2,1).

Any complex optimization problem is subject to overfitting, which is characterized by increasing objective-function values over multiple subsequent optimization iterations~\cite{MontesinosLópez2022}. For the 1D shock,  \emph{a posteriori} optimization  resulted in significant reduction of the loss, though the highly nonlinear dependence of the PDE solution on the parameters---and the nontrivial parameter space dimensionality---generally gave  nonmonotonic loss convergence. To avoid overfitting, we stopped each model's training process early at the first local minimum it encountered; the final parameter state corresponds to that  which gives the smallest loss within the vicinity of the local minimum.
The results in Section~\ref{sec:Results} use the parameter sets corresponding to this ``nonoverfitting minimum'' loss.

\section{Performance of \textit{A Posteriori}-Trained Models} \label{sec:Results}

We now evaluate the five \emph{a posteriori}-optimized parameter sets for the second-order continuum model for predictive accuracy. We compare these simulation results to those of the first-order Navier--Stokes predictions, second-order PP18 predictions, and the target DSMC data.

Table~\ref{tab:Table_parameters} lists the parameter sets of the original PP18 and optimized models. For comparison, we include the \emph{a priori}-trained M8 parameters in Table~\ref{tab:Table_parameters}. Most of the \emph{a priori}-trained parameters have significantly different signs and magnitudes than those of the other models, which is symptomatic of overfitting to the in-sample case, lack of sufficient physical constraints, or both, and underscores the \emph{a priori}-trained model's poor out-of-sample performance. For these reasons, we do not consider the \emph{a priori} parameters further.

\begin{table}[]
\caption{Initial (PP18) and optimized parameter sets.  Unless otherwise indicated, optimized parameter sets used \emph{a posteriori} training.}
\centering
\begin{tabularx}{\textwidth}{p{3.2cm}  *{8}{>{\centering\arraybackslash}X} }
    \toprule
     Model & $k_0^\star$ & $k_0^{\star\star}$ & $\beta_3$ & $\gamma_3$ & $\beta_4$ & $\gamma_4$ & $\beta_5$ & $\gamma_5$ \\ \midrule
    PP18~\cite{paolucci2018second} & 1.000 & 4.000 & 0.500 & 0.730 & 0.500 & 0.500  & 0.500  & 0.500  \\
    \emph{A priori} M8 & -1.353 & 1.659 & 2.965 & 0.391 & -1.761 & -1.758 & -1.830 & -1.830 \\
    M5 & 0.471 & 3.603 & 0.897 & 0.947 & 0.046 & -0.019 & 0.218 & 0.121 \\ 
    M8 & 0.594 & 3.728 & 0.812 & 0.941 & 0.165 & 0.100 & 0.341 & 0.245 \\
    M258~-~$\chi$(1,1,1) & 0.583 & 3.684 & 0.842 & 0.947 & 0.151 & 0.092 & 0.306 & 0.225 \\ 
    M258~-~$\chi$(3,1,1) & 0.561 & 1.076 & 0.595 & 0.997 & -0.280 & 0.282 & -0.813 & -0.197  \\ 
    M258~-~$\chi$(7,2,1) & 0.380 & 2.570 & 0.605 & 1.005 & -0.010 & 0.076 & -0.338 & -0.020  \\ \bottomrule
    \end{tabularx}
\label{tab:Table_parameters}
\end{table}

\begin{figure}
\centering
\includegraphics[width=0.9\linewidth]{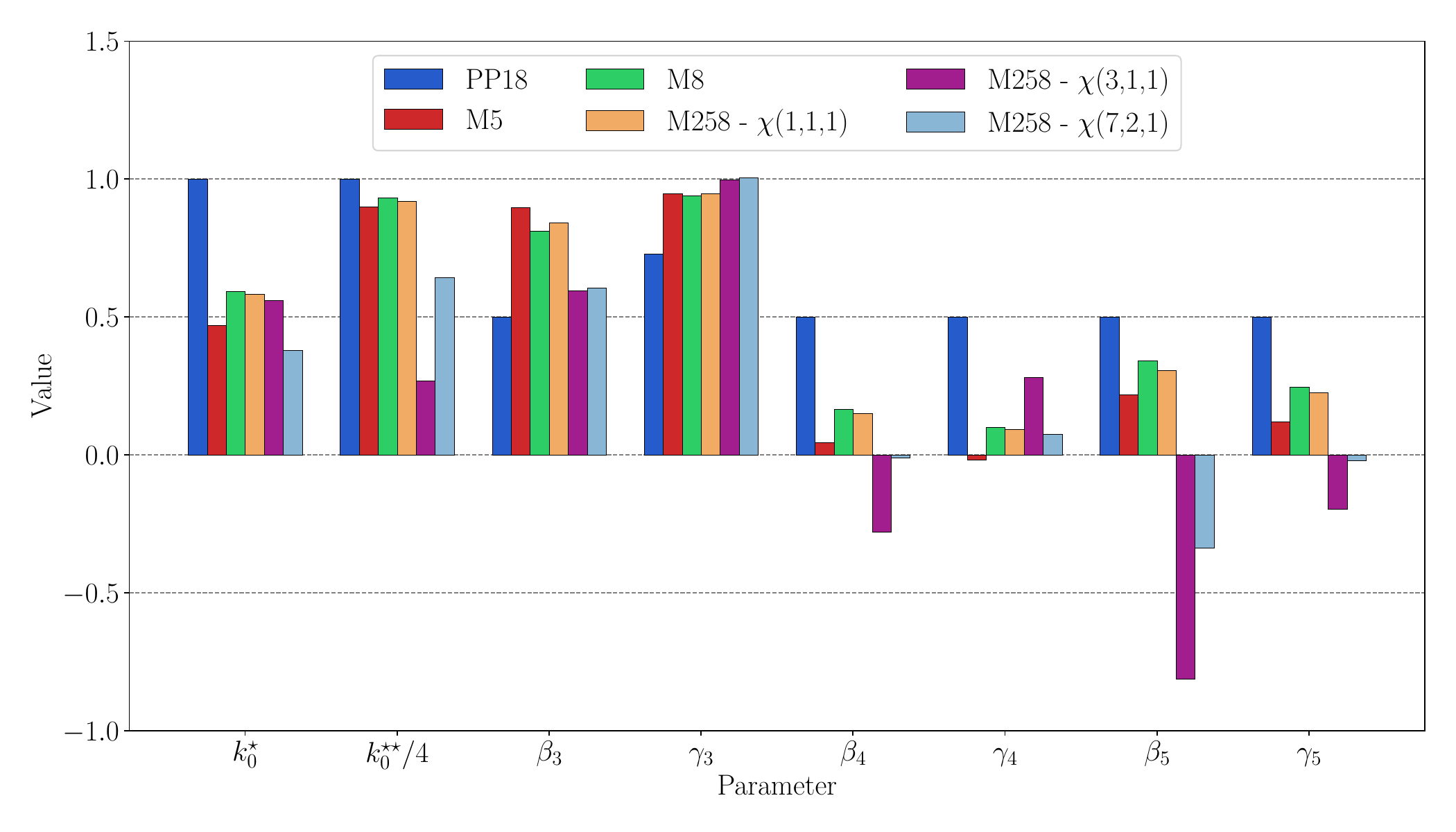}
    \caption{Graphical depiction of initial (PP18) and \textit{a posteriori}-optimized parameter sets.}
\label{fig:Bar_chart}
\end{figure}

Figure~\ref{fig:Bar_chart} displays the parameter values for the initial (PP18) and \emph{a posteriori}-trained models.
In all \emph{a posteriori}-optimized models, the higher-order thermal conductivity power law constants $(k_0^{\star}, k_0^{\star\star})$ decrease from PP18 while remaining positive. The density exponents $(\beta_4, \beta_5)$ are significantly reduced for all optimized models, with negative values for the the bias-trained models. Similarly, the optimized temperature exponents $(\gamma_4, \gamma_5)$ are lower than the PP18 values, with inconclusive sign of those among the optimized models. The optimized first-order thermal conductivity parameters $\beta_3$ and $\gamma_3$ increased from the PP18 values.

For this transition-continuum flow, the parameter changes indicate increased dependence of the first-order thermal conductivity correction factor on temperature gradients and decreased dependence on density gradients. All higher-order thermal conductivities are less sensitive to temperature gradients and more sensitive to density gradients than the PP18 parameters.

\begin{figure}
\centering
\begin{minipage}[t]{0.49\textwidth}
  \centering
  \begin{tikzpicture}[remember picture,overlay]
  \end{tikzpicture}
  \includegraphics[width=\linewidth]{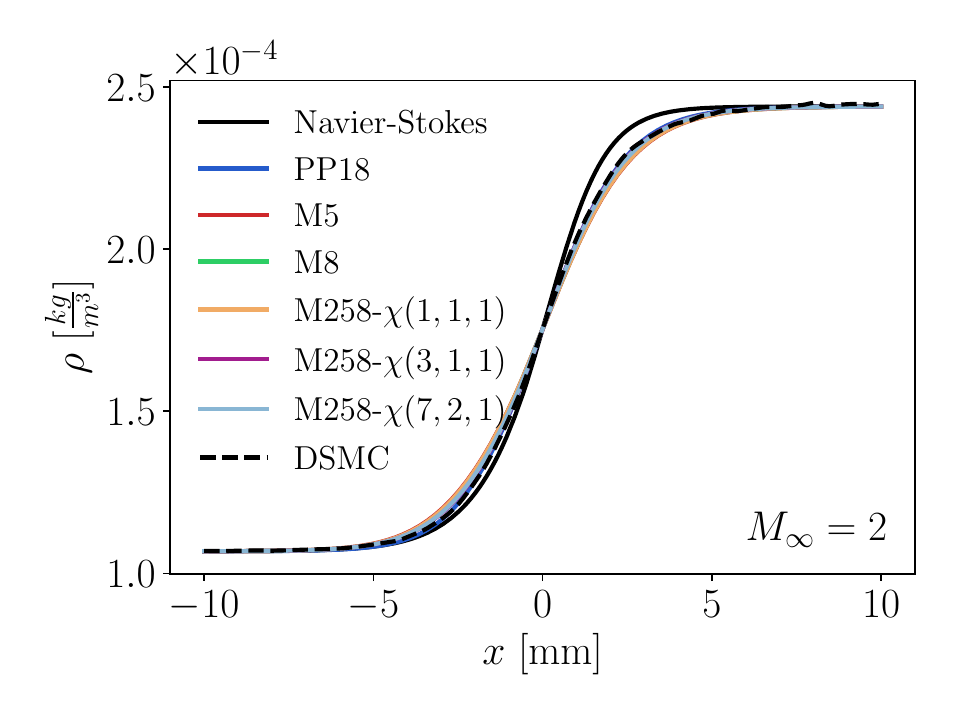}
\end{minipage}
\hfill
\begin{minipage}[t]{0.49\textwidth}
  \centering
  \begin{tikzpicture}[remember picture,overlay]
  \end{tikzpicture}
  \includegraphics[width=\linewidth]{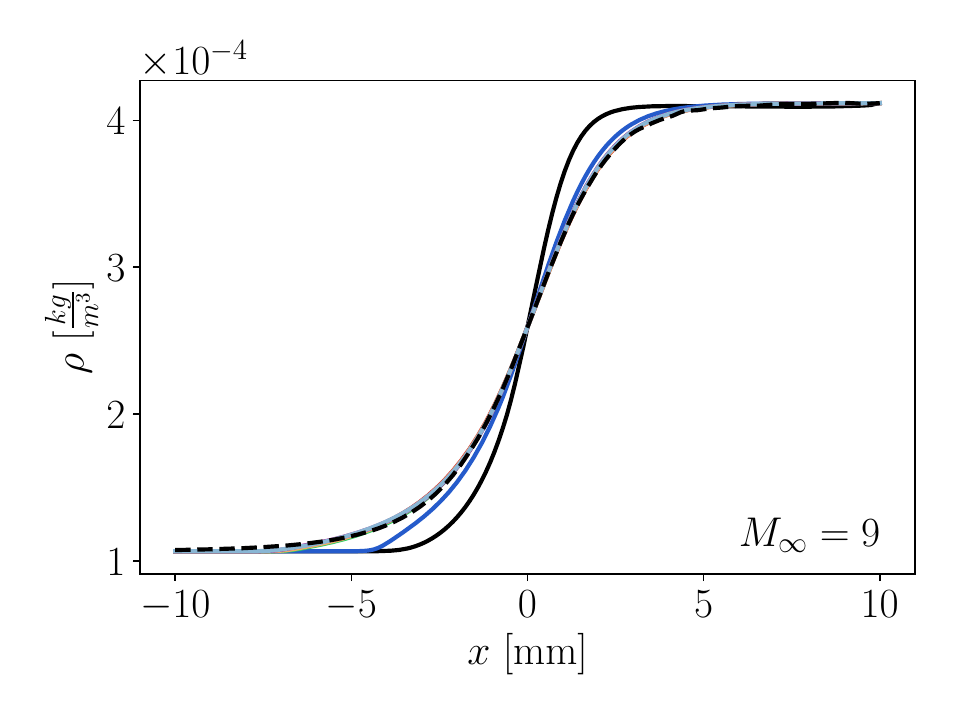}
\end{minipage}
\centering
\begin{minipage}[t]{0.49\textwidth}
  \vspace{-0.2cm}
  \centering
  \begin{tikzpicture}[remember picture,overlay]
  \end{tikzpicture}
  \includegraphics[width=\linewidth]{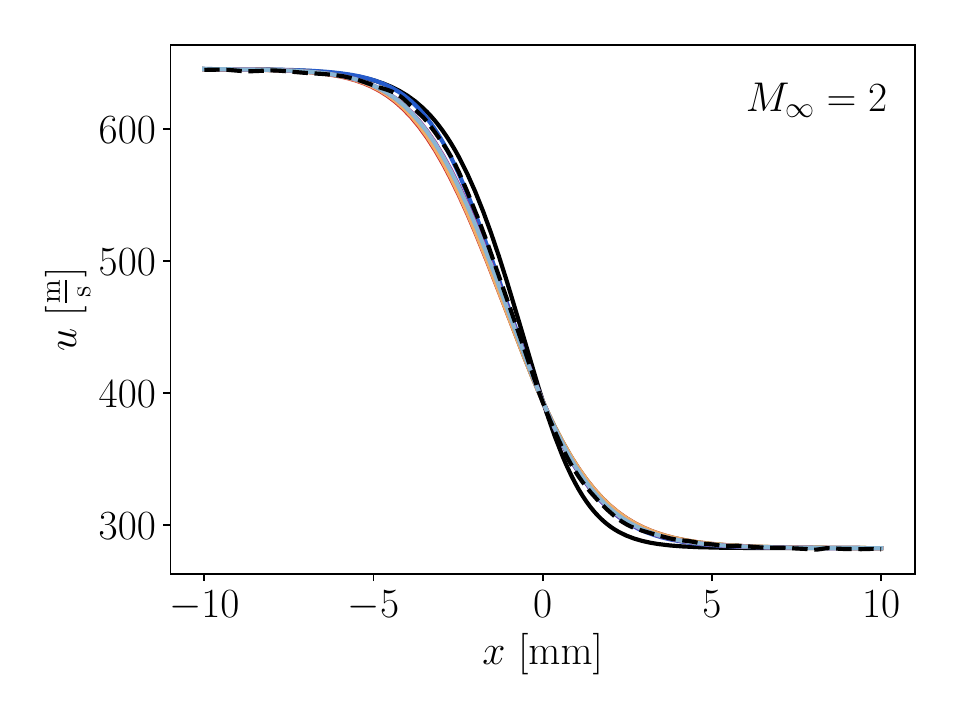}
\end{minipage}
\hfill
\begin{minipage}[t]{0.49\textwidth}
  \vspace{-0.2cm}
  \centering
  \begin{tikzpicture}[remember picture,overlay]
  \end{tikzpicture}
  \includegraphics[width=\linewidth]{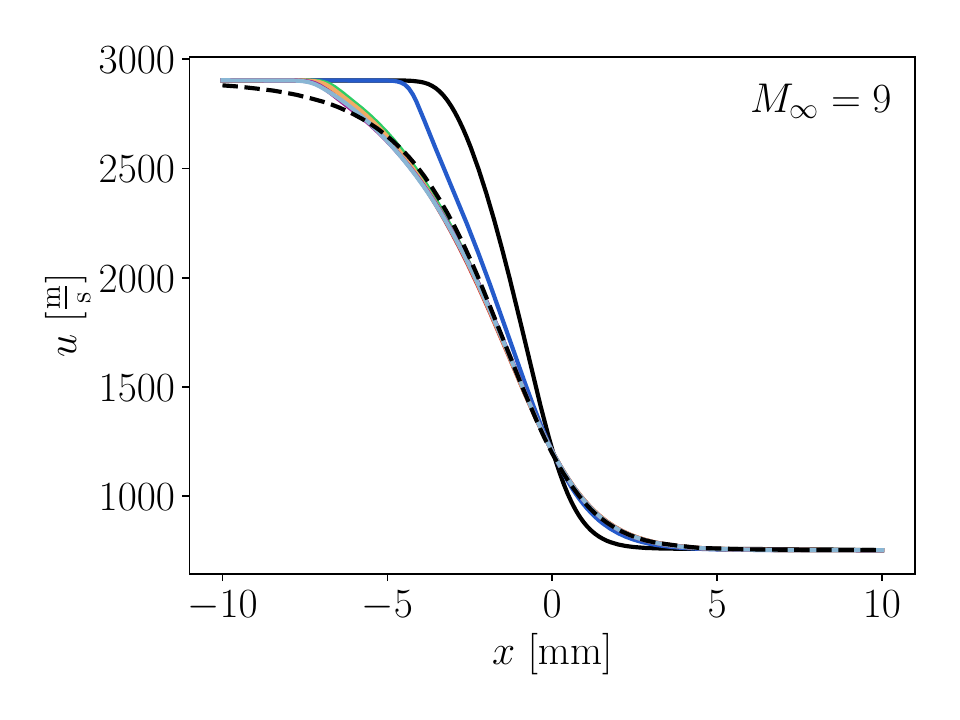}
\end{minipage}
\centering
\begin{minipage}[t]{0.49\textwidth}
  \vspace{-0.2cm}
  \centering
  \begin{tikzpicture}[remember picture,overlay]
  \end{tikzpicture}
  \includegraphics[width=\linewidth]{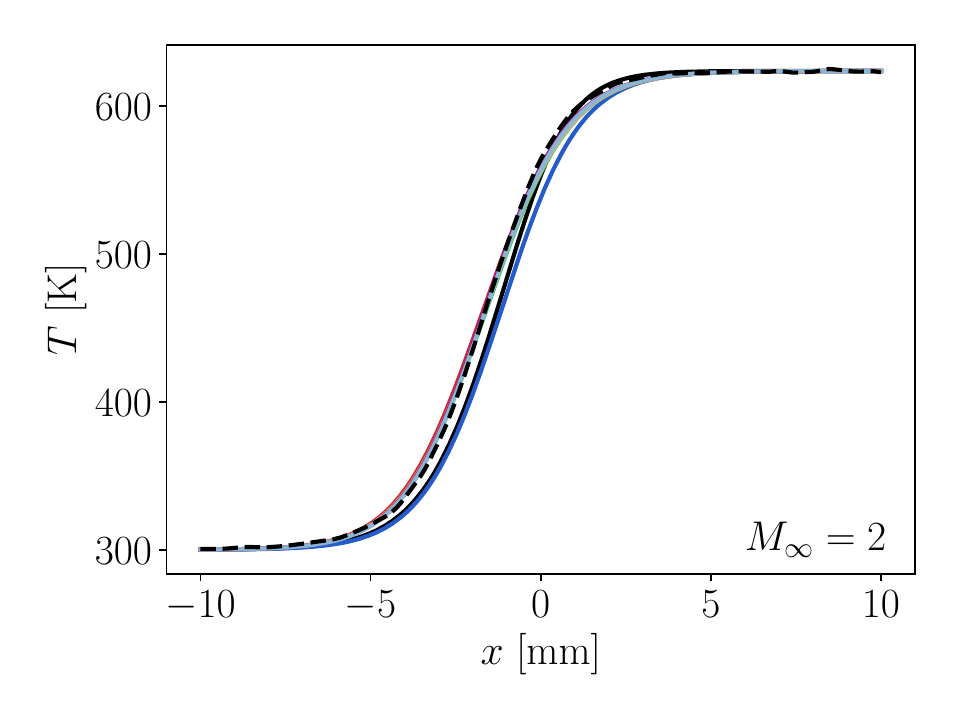}
\end{minipage}
\hfill
\begin{minipage}[t]{0.49\textwidth}
  \centering
  \vspace{-0.2cm}
  \begin{tikzpicture}[remember picture,overlay]
  \end{tikzpicture}
  \includegraphics[width=\linewidth]{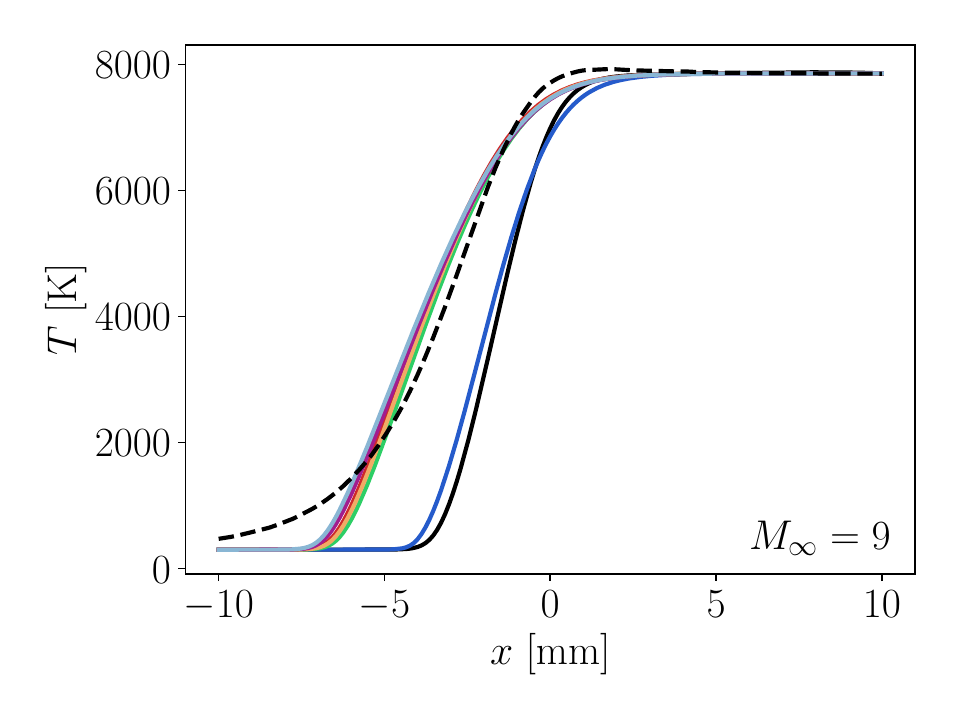}
\end{minipage}
\caption{Density, velocity and temperature profiles for $M_\infty = 2$ (left), and $M_\infty = 9$ (right) normal shocks.}
\label{fig:Aposteriori_grid}
\end{figure}

It is worth noting that the $k^\star$ exponents ($\beta_4$ and $\gamma_4$) and $k^{\star\star}$ temperature exponent ($\gamma_5$) are optimized toward zero from the original PP18 parameters. Shock profiles calculated using a modified M258~-~$\chi$(7,2,1) model with these parameters set explicitly to zero are virtually identical to those using the model parameters presented in this section (see Figure~\ref{fig:Aposteriori_grid}). While it is possible that setting these parameters to zero during training could reduce the potential for overfitting, this is not guaranteed to hold for all flow scenarios---the parameters presented in Table~\ref{tab:Table_parameters} have been optimized for viscous shocks at a single freestream Knudsen number. It is possible that optimized parameters for different flows (e.g., transition-continuum Couette flow) and/or different Knudsen number regimes could have different optimal values of $\beta_4$, $\gamma_4$, and $\gamma_5$ that might not necessarily trend toward zero.

\subsection{Shock Structure}

Figure~\ref{fig:Aposteriori_grid} displays the shock structure (density, velocity and temperature profiles) for $M_\infty = 2$ (quasi-in-sample for M258 models) and $M_\infty = 9$ (out-of-sample for all models).
The first-order Navier--Stokes predictions are inaccurate with respect to the DSMC data at both Mach numbers, with particularly large deviations for $M_\infty=9$. The PP18 second-order model is accurate for the $M_\infty=2$ density and velocity profiles, but suboptimal predictions of the heat flux (see Section~\ref{sec:heatflux}) result in an inaccurate upstream temperature profile. The PP18 model predicts a remarkably accurate density profile at $M_\infty=9$ but has an unphysical ``kink'' in the velocity profile, and its temperature prediction is relatively unaffected from the first-order solution due to its inaccurate heat flux. All optimized models give solutions that capture the shock geometry and match the DSMC profiles, with the M258~-~$\chi$(3,1,1) and M258~-~$\chi$(7,2,1) models providing the most accurate predictions. The parameter optimization process mitigates the PP18 parameters' deficiencies, particularly at high Mach numbers. 

\begin{figure}
    \centering
    \begin{minipage}[t]{0.49\textwidth}
        \centering
        \includegraphics[width=\linewidth]{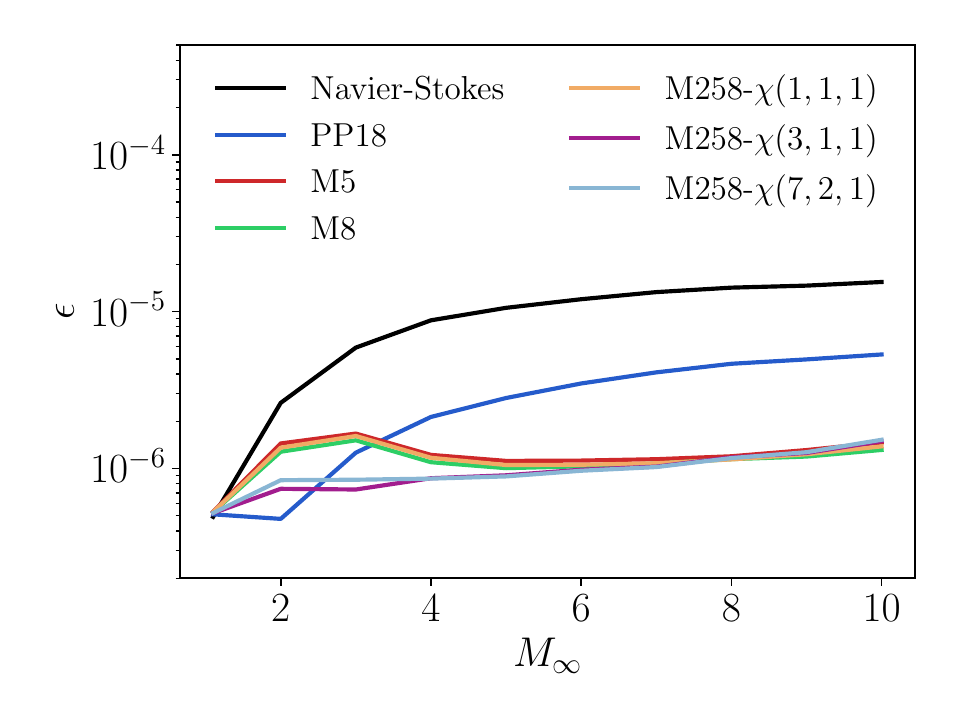}
        \begin{tikzpicture}[remember picture,overlay]
            \node[anchor=north west] at (-3.5,6.25) {\textbf{a)}};
        \end{tikzpicture}
    \end{minipage}
    \hfill
    \begin{minipage}[t]{0.49\textwidth}
        \centering
        \includegraphics[width=\linewidth]{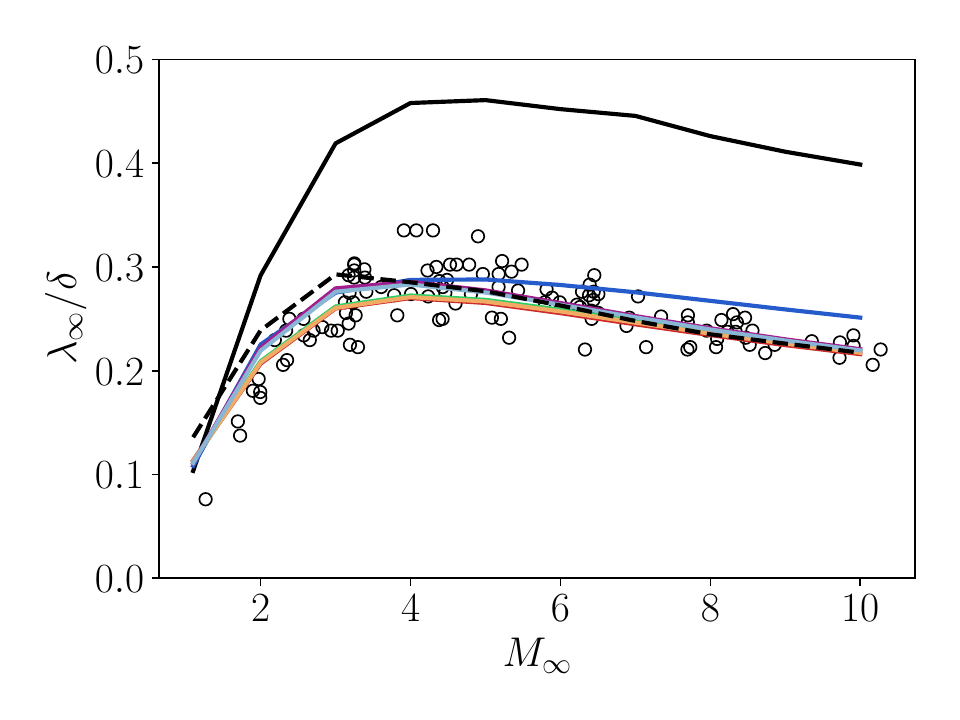}
        \begin{tikzpicture}[remember picture,overlay]
            \node[anchor=north west] at (-3.95,6.25) {\textbf{b)}};
        \end{tikzpicture}
    \end{minipage}
    \vspace{-0.5cm}
    \caption{Density error (in $\mathrm{kg}/\mathrm{m}^3$) (a) and the inverse shock thickness (b) for first-order Navier--Stokes, second-order PP18, and \textit{a posteriori}-optimized second-order models across the range of tested Mach numbers.}
    \label{fig:RhoErr_ShockThick}
\end{figure}

We now consider the full-domain prediction errors across the range of tested Mach numbers. Figure~\ref{fig:RhoErr_ShockThick}(a) plots the L2 density error $(\epsilon)$ of the first-order Navier--Stokes, second-order PP18, and optimized second-order predictions against the tested $M_\infty$. The optimized models significantly improve the density-profile predictions at the higher Mach numbers, and the bias-optimized models are significantly more accurate than the unbiased models at the lower Mach numbers.

The inverse shock thickness provides another accuracy metric and has been reported in numerous experimental studies~\cite{Alsmeyer_1976, 10.1063/1.1711007, doi:10.2514/6.1964-35, Schmidt_1969, doi:10.2514/3.49425}, though it is not as comprehensive as full-domain errors. We calculate the shock thickness as $\delta=( \rho_2-\rho_\infty)/\max{( \partial \rho/\partial x)}$, where $\rho_2$ is the post-shock density. Figure~\ref{fig:RhoErr_ShockThick}(b) shows the inverse shock thickness normalized by the upstream mean free path ($\lambda_\infty = 1.098$~mm). All second-order models vastly outperform the first-order Navier--Stokes model, though the unbiased (M5, M8 and M258~-~$\chi$(1,1,1)) parameter sets perform worse than the PP18 parameters, relative to the DSMC data, for $M_{\infty} \leq 3$, though all are within the range of the experimental uncertainty. On the contrary, the biased models (M258~-~$\chi$(3,1,1) and M258~-~$\chi$(7,2,1)) are equally accurate to PP18 for $M_{\infty} \leq 3$. All optimized models are more accurate than both the first-order and PP18 models for $M_\infty>4$. The model biases introduced for minibatch-trained models improve the overall model accuracy across the entire Mach number range, which demonstrates the effectiveness of the approach.

We also analyze the shock asymmetry and temperature--density separation distance. The shock asymmetry quotient is defined as
\begin{equation*}
Q = \frac{\int_{-\infty}^{x_{0,\rho}} \rho_{*} dx}{\int_{x_{0,\rho}}^{\infty} 1 - \rho_{*} dx},
\end{equation*}
where $\rho_{*}(x) = \left(\rho(x) - \rho_{\infty}\right)/\left(\rho_2 - \rho_{\infty}\right)$ is the normalized density, and $\rho_2$ is the downstream density. The density midpoint $x_{0,\rho}$ is obtained as $x_{0,\rho}\ |\ \rho_{*}(x_{0,\rho}) = 0.5$. The temperature--density separation distance is
\begin{equation*}
  \Delta_{\rho T} = \frac{1}{\lambda_\infty}(x_{0,\rho} - x_{0,T}),
\end{equation*}
where $x_{0,T}$ is the temperature midpoint, defined analogously to the density midpoint using a normalized temperature $T_{*} = \left(T(x) - T_{\infty}\right)/\left(T_2 - T_{\infty}\right)$.

\begin{figure}
    \begin{minipage}[t]{0.49\textwidth}
        \centering
        \vspace{-0.2cm}
        \includegraphics[width=\linewidth]{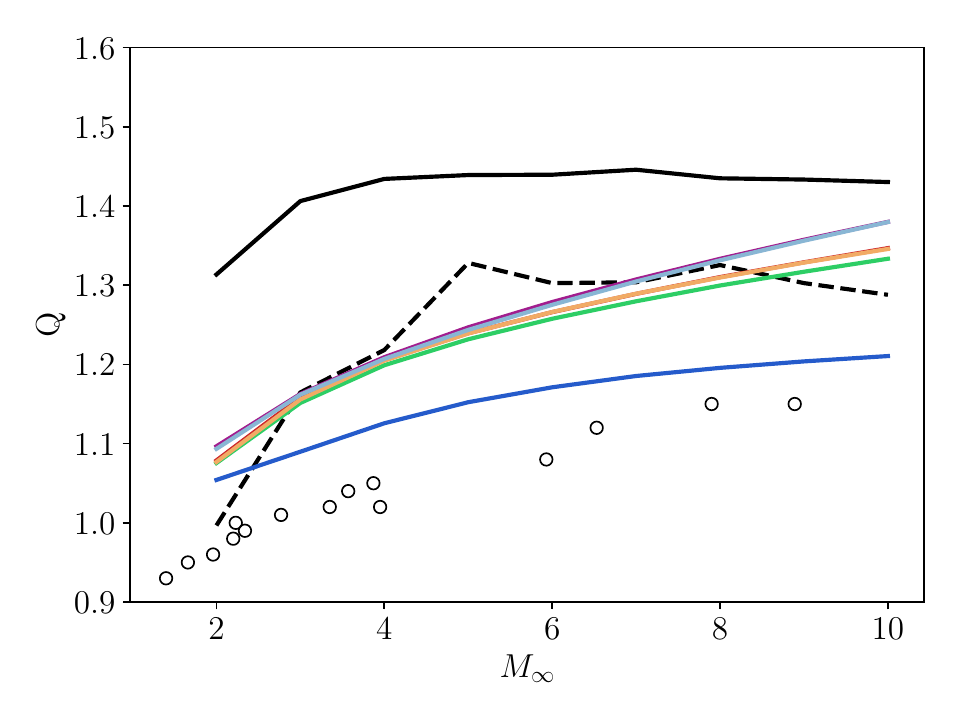}
    \end{minipage}
    \hfill
    \begin{minipage}[t]{0.49\textwidth}
        \centering
        \vspace{-0.2cm}
        \includegraphics[width=\linewidth]{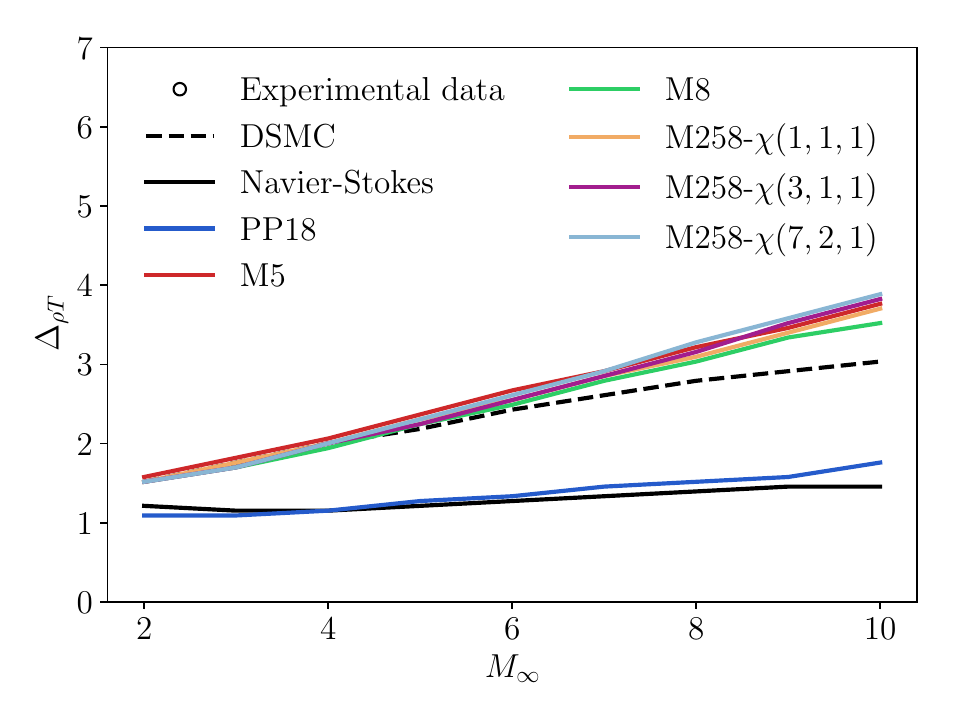}
        \end{minipage}
    \caption{Shock asymmetry quotient (left) and temperature--density separation distance (right) for $M_{\infty}\in[2, 10]$.}
    \label{fig:Asymmetry_and_rho_T_distance}
\end{figure}

Figure~\ref{fig:Asymmetry_and_rho_T_distance} shows the shock asymmetry and temperature--density separation distance calculated from the DSMC target data, standard first-order Navier--Stokes, second-order PP18 Navier--Stokes, and \emph{a posteriori}-optimized second-order models.
The optimized second-order models significantly improve the agreement of the simulated shock asymmetry with the DSMC target data, as could be anticipated by these models' improvements with respect to the DSMC in other metrics. The figure also includes experimental shock asymmetry data \cite{Schmidt_1969,Alsmeyer_1976}, with which the PP18 solutions agree better than the \emph{a posteriori}-optimized and first-order solutions, though it is important to note that these experimental data points were not included in the target data for optimization. Whether the DSMC solutions should agree these experimental data points is beyond the scope of the present manuscript.

Likewise, the temperature--density separation distance of the optimized second-order solutions shows excellent agreement with the DSMC target data. Both the standard Navier--Stokes and the second-order PP18 model predictions significantly underestimate the temperature--density separation over the range of Mach numbers. The optimized second-order models also capture the slope of the temperature--density separation distance with Mach number, which is not well predicted by the first-order and second-order PP18 Navier--Stokes solutions.

\begin{figure}
\centering
\begin{minipage}[t]{0.49\textwidth}
  \centering
  \vspace{-0.2cm}
  \includegraphics[width=\linewidth]{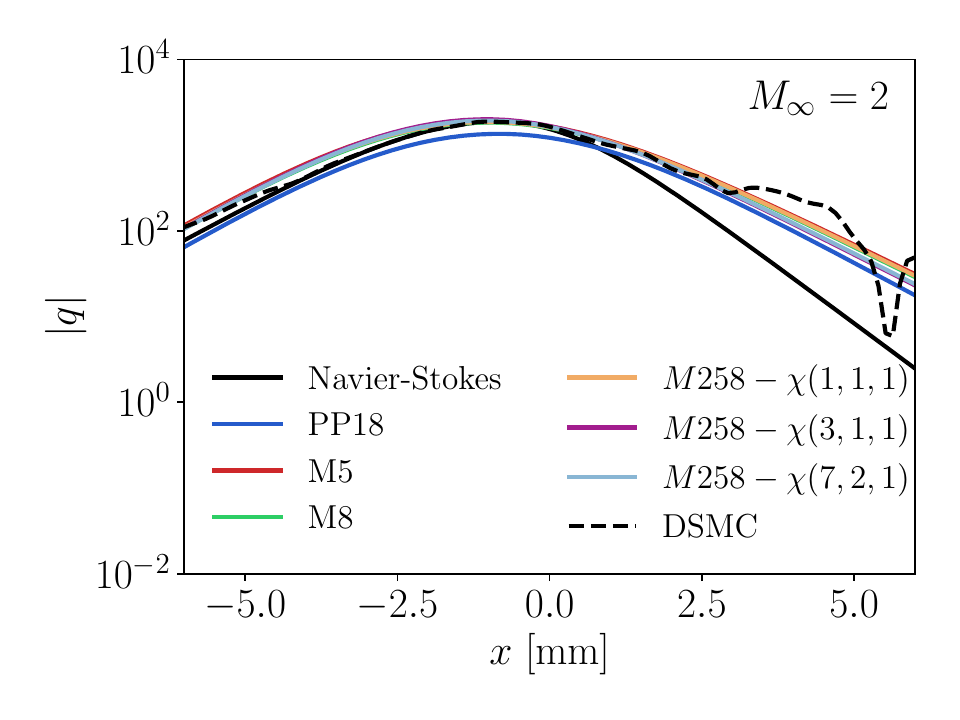}
\end{minipage}
\hfill
\begin{minipage}[t]{0.49\textwidth}
  \centering
  \vspace{-0.2cm}
  \includegraphics[width=\linewidth]{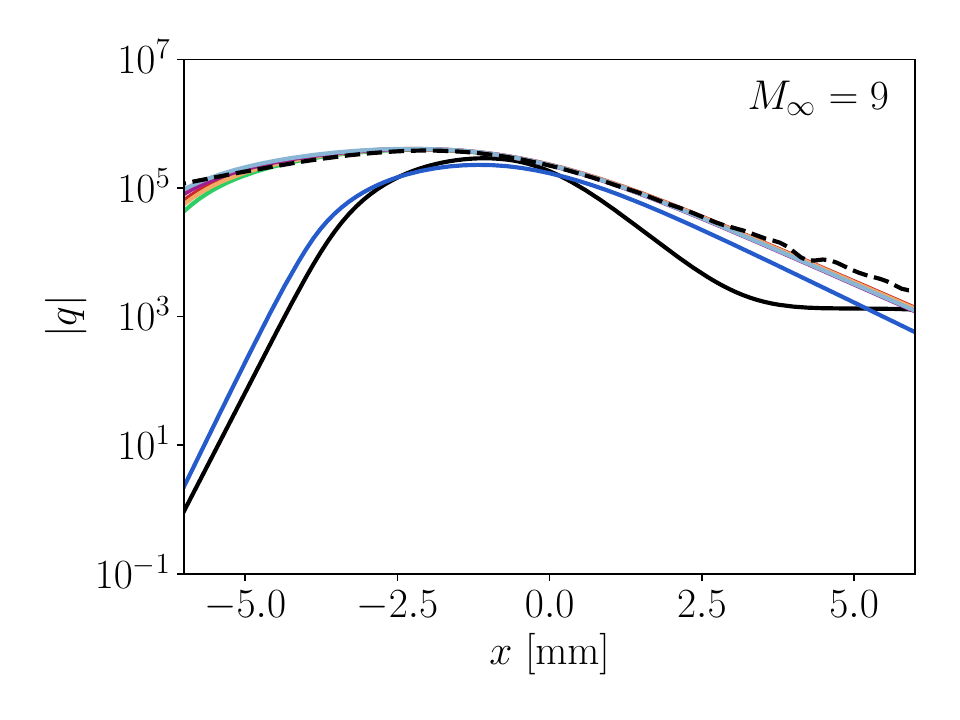}
\end{minipage}
\begin{minipage}[t]{0.49\textwidth}
  \vspace{-0.2cm}
  \centering
  \includegraphics[width=\linewidth]{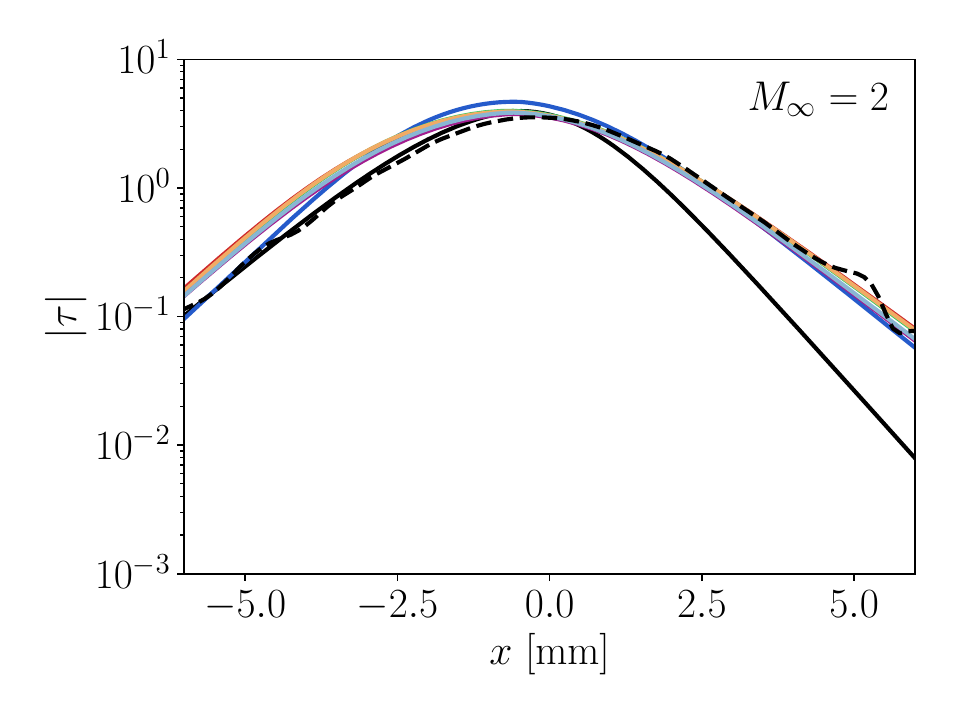}
\end{minipage}
\hfill
\begin{minipage}[t]{0.49\textwidth}
  \vspace{-0.2cm}
  \centering
  \includegraphics[width=\linewidth]{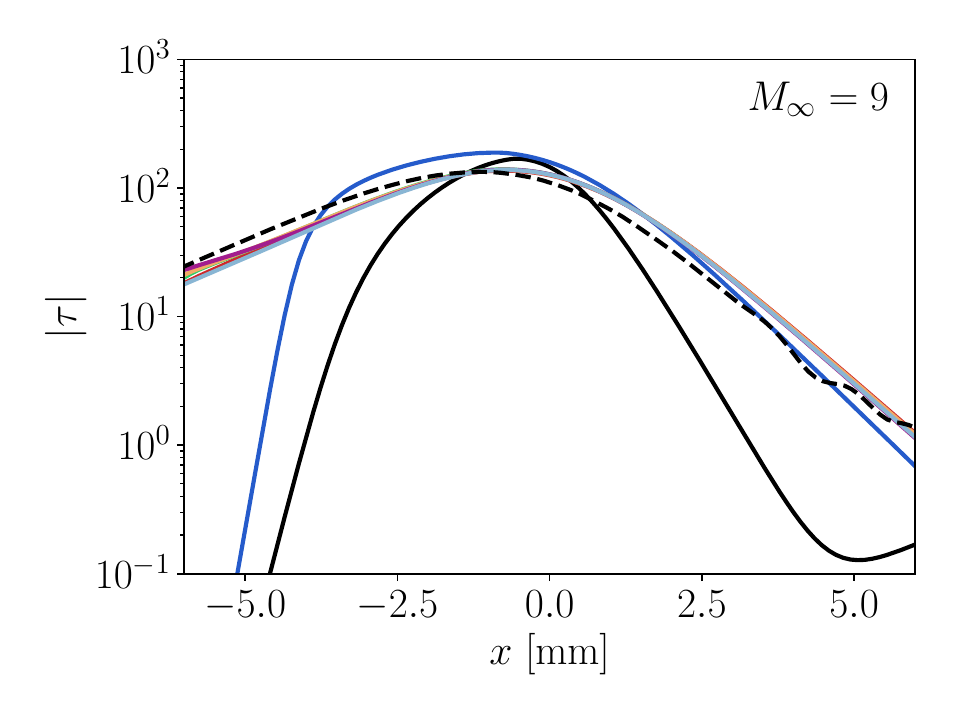}
\end{minipage}
\caption{Heat flux (top) and viscous stress (bottom) magnitude for $M_\infty = 2$ and $M_\infty = 9$ shocks produced by the first-order, PP18 second-order, and \textit{a posteriori}-optimized second-order models.}
\label{fig:tau_q}
\end{figure}

\subsection{Viscous Stress and Heat Flux}\label{sec:heatflux}

One concern with PDE-embedded optimization is that the objective function (the error of the PDE solution) may be minimized by introducing unphysical closures---the model could be ``right for the wrong reasons''---which could be considered a form of physical overfitting. Unlike for \textit{a priori} training (Section~\ref{sec:a_priori}), the \textit{a posteriori} training process does not include the viscous stress and heat flux in its loss function \eqref{eq:loss}, yet the optimized models still predict these quantities. The similarity of the modeled viscous stress and heat flux to those obtained from the Boltzmann distribution function can be used to assess the physical realism of the optimized second-order models.

Examination of viscous stress and heat flux at $M_{\infty} = 2$ and $M_{\infty} = 9$, shown in Figure~\ref{fig:tau_q}, demonstrates extremely close agreement of the optimized models' predictions with the DSMC data, with significant improvement over the first-order Navier--Stokes--Fourier and PP18 second-order models. The results confirm that the optimized models capture the true physics of these nonequilibrium shocks. The performance of the optimized models for other upstream Mach numbers is similar.

\section{Conclusion}~\label{sec:Summary}
We apply PDE-constrained optimization to obtain the parameters of a second-order continuum transport model and evaluate the trained models for transition-continuum shocks in argon. The second-order continuum theory provides mathematical descriptions of the heat flux and stress tensor that depend on second-order temperature and density gradients as well as invariants of the strain-rate tensor. The PDE-constrained optimization method, based on the adjoints of the flow variables, is able to optimize objective functions comprising the PDE solutions, not necessarily the higher-order terms themselves, which is important in the present case for which such terms would be challenging to obtain experimentally or from a higher-order (e.g., Boltzmann equation) solution. While of secondary importance for the present models, the adjoint-based method is also able to optimize efficiently over high-dimensional parameter spaces, as would be the case for deep learning models, for example.  We obtain target data, used in the optimization objective functions, by integrating DSMC solutions of the Boltzmann equation for the same flows. 

We evaluate the fidelity of the optimized second-order models by comparing their predicted density, velocity, and temperature fields to those obtained from the DSMC targets. In all cases, the optimization method is able to improve the predictive accuracy from both the first-order Navier--Stokes equations and the second-order continuum model with its originally proposed parameters.  The physical validity of the optimized models was examined by comparing their predicted heat flux and viscous stress profiles, not included in the objective function, to those obtained from the target Boltzmann equation solutions. The extremely close match of the optimized models to these transport terms indicates that the learned models correctly capture the nonequilibrium physics of the problem.

We evaluate the out-of-sample solution accuracy of the \textit{a posteriori}-trained models by testing over a range of Mach numbers, both interpolating between those  used for optimization and extrapolating outside the training range. We introduce an objective function biasing technique to avoid overfitting to training conditions with initially larger errors (e.g., higher Mach numbers), which increases model accuracy for minibatch-trained  cases without negatively impacting the solution accuracy of the initially higher-error cases.

We additionally compare the PDE-constrained, \emph{a posteriori} approach to a PDE-decoupled, \textit{a priori} optimization technique commonly used in scientific machine learning. While the models obtained through \textit{a posteriori} optimization generate stable shock solutions with significantly improved accuracy, models obtained through  the \textit{a priori} approach suffer numerical instability when used for predictions, which results in poor out-of-sample accuracy, despite the models' satisfactory convergence to the DSMC target profiles during optimization. This highlights the importance of PDE-constrained optimization for highly nonlinear systems such as hypersonic transition-continuum flows.

The present work confirms that the second-order continuum theory of Paolucci \& Paolucci can be optimized for predictive accuracy across a range of nonequilibrium transition-continuum conditions, at which it achieves comparable accuracy to DSMC at slightly greater cost than the standard first-order continuum models. The second-order continuum model in two and three space dimensions has several additional parameters, which the adjoint-based method is expected to be capable of optimizing as well.
We are currently optimizing the second-order continuum theory for these flows with the goal of obtaining accurate, generalizable, predictive transport models for the transition-continuum regime.

\section*{Acknowledgments}
This work was supported by the Office of Naval Research under award N00014-22-1-2441. The authors are grateful to Prof.~Iain Boyd for sharing his DSMC solver.  The authors acknowledge computational time on resources supported by the University of Notre Dame Center for Research Computing.


\end{document}